\documentclass[preprint,aps,prd,groupedaddress,showpacs,nofootinbib]{revtex4-1}%
\usepackage{graphicx}
\usepackage{amsmath}
\usepackage{amssymb}
\usepackage{bm}
\usepackage{epsfig}
\usepackage{dcolumn}
\usepackage{slashed}
\usepackage{empheq}

\newcommand{\bea}{\begin{eqnarray}}
\newcommand{\eea}{\end{eqnarray}}
\newcommand{\nn}{\nonumber}

\begin{document}

\setlength\baselineskip{17pt}

\title{NNLO soft function for threshold single inclusive jet production}

\author{Yaroslav Balytskyi}
\email{ybalytsk@uccs.edu}
\affiliation{Department of Physics and Energy Science, University of Colorado at Colorado Springs, Colorado Springs, CO 80933, USA}

\author{Jianbo Gao}
\email{jianbo.gao@ceamat.com}
\affiliation{Centre of Excellence for Advanced Materials, Dongguan 523808, China}

\date{\today}

\begin{abstract}
We present a computation of the global soft function for single-inclusive hadronic jet production at next-to-next-to-leading order (NNLO) in the strong coupling constant involving four light-like Wilson lines. Our soft function belongs to SCET$_{\text I}$ observables obeying the non-Abelian exponentiation theorem. We provide the calculation of the soft function involving four collinear light-cone directions  for all $2\rightarrow2$ processes extending the previous result. We perform the threshold resummation for these processes at the next-to-next-to-leading logarithmic accuracy and present the corresponding numerical results.

\end{abstract}

\maketitle

\section{Introduction}
The new data obtained with an increasingly high precision at the Large Hadron Collider (LHC) and other experiments provides an opportunity to search for new physics from a small mismatch between the Standard Model (SM) theoretical predictions and experimental measurements. The only available theoretical tool for making such predictions from first principles is the Quantum Chromodynamics (QCD) perturbative expansion in the strong coupling constant $\alpha_S$ or $\alpha_SL$ if large logarithm $L$ of fraction of scales is present. To reliably interpret the collider data, the precision of QCD theoretical predictions has to match the small experimental errors. Such calculations are carried out at the fixed-order in QCD perturbation theory and may include the resummation of large logarithms in particular regions of phase space, by analytic resummation or through the use of parton-shower simulations. These days, considerable efforts have been made towards the development of theoretical predictions for physical processes at colliders with a complete control over the final state kinematics beyond the next-to-leading order (NLO) accuracy. With regards to fixed-order calculations, over the past few years, significant results have been achieved. These include the calculations of fully differential gluon-gluon fusion Higgs production~\cite{Overview_1, Overview_2}, the single jet production in deep-inelastic scattering~\cite{Overview_3} at $N^3LO$ in the strong coupling constant, the NNLO calculations of the single inclusive jet or dijet production~\cite{Overview_4}, $V/N + j$ processes~\cite{Overview_5, Overview_6, Overview_7, Overview_8, Overview_9, Overview_10, Overview_11}, Higgs production through the vector boson fusion~\cite{Overview_12, Overview_13}, single top~\cite{Overview_14, Overview_15} and $t\bar{t}$ production~\cite{Overview_16, Overview_17}  at the LHC, the jet productions in deep-inelastic scattering~\cite{Overview_18, Overview_19}, and the heavy flavor production in deep-inelastic neutrino scattering~\cite{Overview_20}. The developments in the higher order calculations rely on the new higher loop calculations as well as on the new theoretical schemes for treating the singularities in the real emissions. The current status of the loop calculations for the collider processes can be illustrated by the calculation of the analytic two-loop five-point amplitude~\cite{Overview_21, Overview_22, Overview_23}, as well as by the new ideas that appear~\cite{Overview_24, Overview_25, Overview_26, Overview_27}.

Jets are crucial at high-energy particle and nuclear physics frontiers serving as a tool for precision tests of the SM as well as for the searches for new physics at the TeV scale at the LHC. As demonstrated by the measurements at the LHC and the Relativistic Heavy Ion Collider (RHIC), they are also a unique probes of non-perturbative dynamics, such as collinear parton distribution functions (PDFs)~\cite{NOverview_1} and transverse momentum dependent ones (TMDPDFs)~\cite{NOverview_2}, the intrinsic spin of the nucleon~\cite{NOverview_3,NOverview_4,NOverview_5} and the hot medium effects in the quark-gluon plasma~\cite{NOverview_6,NOverview_7,NOverview_8}. The experimental advances are supplemented by the progress in development of new jet substructure techniques~\cite{NOverview_9,NOverview_10}, accurate extractions of the SM parameters~\cite{NOverview_10,NOverview_11} as well as the three-dimensional tomographic images of nucleons ~\cite{NOverview_12,NOverview_13,NOverview_14,NOverview_15,NOverview_16,NOverview_18} out of jets. These developments will receive additional impetus at the future Electron-Ion Collider (EIC)~\cite{NOverview_19,NOverview_20},
additionally enabling the possibility to consider the polarization degrees of freedom.

The double differential cross section for $pp\rightarrow$ jet $+ X$ process has the following form:

$$
\frac{p^2_T\mathrm{d}^2\sigma }{\mathrm{d}p^2_T \mathrm{d}\eta} = \sum_{i_1,i_2}\int^{V\left(1- W\right)}_0 \mathrm{d}z \int^{1 - \frac{1 - V}{1 - z}}_{\frac{VW}{1 - z}} \mathrm{d}v\ x_1^2 f_{i_1}\left(x_1\right) x_2^2 f_{i_2}\left(x_2\right)\frac{\mathrm{d}^2\hat{\sigma}_{i_1i_2}}{\mathrm{d}v\mathrm{d}z}\left(v, z, p_T, R\right)
$$

with $p_T$ being the transverse momentum and $\eta$ the rapidity of the signal jet, $V = 1 - \frac{p_Te^{-\eta}}{\sqrt{S}}$, $VW = \frac{p_Te^{-\eta}}{\sqrt{S}}$, and the hadronic center-of-mass (CM) energy is $\sqrt{S}$. The PDFs $f_i$ depend on the momentum fractions $x_1 = \frac{VW}{v\left(1 - z\right)}$ and $x_2 = \frac{1 - V}{\left(1 - v\right)\left(1 - z\right)}$, and the sum is taken over all partonic channels initiating the process with the cross sections given by $\hat{\sigma}_{i_1i_2}$. The kinematic variables are defined as:
$t = \left(p_1 - p_3\right)^2$ and $u = \left(p_2 - p_3\right)^2$ with $p_{1,2}$ being the momenta of the two incoming partons, $p_3$ is the momentum of parton which initiates the signal jet, and $p_4$ is the momentum of the recoiling system against the signal jet. The partonic cross sections $\hat{\sigma}_{i_1i_2}$ are  functions of $p_T$ and partonic kinematic variables $s = x_1x_2S$, $v = \frac{u}{u + t}$ and $z$. They can be further factorized in the $R\rightarrow0$ and $z\rightarrow0$ limit:

$$
\frac{\mathrm{d}^2\hat{\sigma}_{i_1i_2}}{\mathrm{d}v\mathrm{d}z} = s\int\mathrm{d}s_X\mathrm{d}s_c\mathrm{d}s_G\delta\left(zs-s_X-s_G-s_c\right)Tr\left[H_{i_1i_2}\left(v,p_T,\mu_h,\mu\right)S_G\left(s_G,\mu_{s_G},\mu\right)\right]\times$$
$$
\times J_X\left(s_X,\mu_X,\mu\right)\sum_m Tr\left[J_m\left(p_TR,\mu_J,\mu\right)\otimes_\Omega S_{c,m}\left(s_cR, \mu_{sc},\mu\right)\right]
$$

with the traces taken in color space. The sum is taken taken over all collinear splittings with $m$ being the energetic collinear parton multiplicity, and the associated angular integrals are denoted by $\otimes_\Omega$~\cite{NOverview_21}. The jet is constructed by the anti-$k_T$ algorithm~\cite{NOverview_22}, and the finite mass of the signal jet is allowed. Resummation is performed by solving the renormalization group (RG) equation in Soft-Collinear Effective Theory (SCET) ~\cite{SCET_1,SCET_2,SCET_3, SCET_4, SCET_5, SCET_6}. The hard function $H$ describes the effect of hard radiation and is known to NNLO for numerous phenomenologically interesting processes containing final-state jets ~\cite{Hard_1, Hard_2, Hard_3, Hard_4, Hard_5, Hard_6, Hard_7}. The jet function describes the final jet collinear radiation, and recoil jet function $J_X$ is known to three loops both for the quark and gluon jets~\cite{Jet_1, Jet_2} while the signal jet function $J_m$ is known to 2-loop for the quark jet~\cite{JetXL}. The global soft function $S_G$ is the focus of our paper. It accounts for the wide-angle soft radiation which cannot resolve the small radius $R$ of the jet.

Threshold logarithms play a role near the exclusive phase space boundary when the production of the signal-jet becomes possible and the invariant mass $\sqrt{s_4}$ of the recoiling jet vanishes. The signal jet, however, has a finite invariant mass at threshold, and the radiation inside the jet cone is possible. After the cancellation of infrared divergences, the logarithms of the form $\alpha_S^n\left(\ln^k\left(z\right)/z\right)_+$ remain, with $k \le 2n - 1$, $z = s_4/s$, and $s$ being the  partonic center-of-mass energy. These terms are large in the threshold limit $z\rightarrow 0$ and have to be resummed to all orders in order to obtain reliable perturbative results. Since the parton luminosity functions are steeply falling, these terms dominate over a wide range of the jet-$p_T$ even far from the hadronic threshold as shown in~\cite{PRL_5}.

The global soft function $S_G$ which is the focus of our paper is defined in terms of Wilson lines along the directions $n_i^{\mu} = \frac{p_i^{\mu}}{E_i}$ of the partons participating in the hard scattering. With the use of the color-space formalism ~\cite{Color_1, Color_2}, Wilson line for a particle in a color representation with generators $T^a_i$ is defined as a path ordered exponential:
$$
S_i\left(x\right) =  \mathcal{P}exp\left(ig_S\int^0_{-\infty} \mathrm{d} t \ n_i\cdot A^a\left(x + tn_i\right)T^a_i\right)
$$

For a gluon carrying a color index $c$, the generator is $\left(T^a\right)_{bc} = -if_{abc}$. For an outgoing quark (or incoming anti-quark) with index $\nu$, the generator is $\left(T^a\right)_{\mu\nu} = t^a_{\mu\nu}$ and for an incoming quark (or outgoing anti-quark)  $\left(T^a\right)_{\mu\nu} = -t^a_{\nu\mu}$ respectively. The soft function measures the probability of the soft emissions from $N$ Wilson lines with the momentum component $\omega = n_N\cdot p_X$ along the recoiling jet direction and is defined as:
$$
S = \sum_X \delta(\omega - \sum_i n_N \cdot k_i ) \langle0| \left(S_{1}S_{2} \cdots S_{N}\right)^\dagger|X\rangle\langle X|\left(S_{1}S_{2} \cdots S_{N}\right)|0\rangle
$$

where $S_i \equiv S_i\left(0\right)$. Our soft function belongs to SCET$_{\text I}$ observables which obey the non-Abelian exponentiation (NAE) theorem ~\cite{NAE1, NAE2}. At NNLO for the case of three collinear directions, $N = 3$, the soft function was calculated in~\cite{Becher2012}. The aim of our paper is to extend this approach for the case of four collinear directions, $N = 4$, and calculate the global soft function for a  $ p_1 p_2 \to p_3 p_4$ process with $p_i^2 = 0$ to the next-to-next-to-leading order in $\alpha_s$.  For the case of four collinear directions, the color structure is not diagonal as for the case of three collinear directions. For each color configuration, we calculate the corresponding matrix elements and perform the resummation of the large logarithms. Our manuscript is organized as follows.  In Section~\ref{sec:Derivation}, we provide the derivation and the analytical results on the soft function involving four light-cone directions. In Section~\ref{sec:Renorm_Resum}, we perform renormalization and resummation. In  Section~\ref{sec:Numerical_Results}, we provide the corresponding numerical results, and conclude in Section~\ref{sec:Conclusions}. Auxiliary relations are provided in Section~\ref{Appendices}.

\section{Derivation}\label{sec:Derivation}
Unlike~\cite{Becher2012}, instead of calculating the soft function diagram by diagram, we use the corresponding matrix elements. The formal form of the global soft function for $i$-real-emissions is given by:
\bea
S_G(\omega) =  \int \prod_i \frac{\mathrm{d}^d k_i}{(2\pi)^{d-1}} \delta(k_i^2) \theta(k_i^0) 
\, |{\cal M}|^2_{eik.}[\{k_i\}] \,
\delta(\omega - \sum_i n_4 \cdot k_i ) \,,
\eea
where $n_4$ is the light-like vector along the momentum $p_4$, $p_4 = E_4 n_4$, and we have assumed that the eikonal matrix element $|{\cal M}|^2_{eik.}$ has already included the virtual corrections.

For the later use, we employ the notation:
\bea
[\mathrm{d} k_i] \equiv \frac{\mathrm{d}^d k_i}{(2\pi)^{d-1}} \delta(k_i^2) \theta(k_i^0) 
\eea
Our situation is similar to ~\cite{Becher2012} with the only difference that in ~\cite{Becher2012} there are only $3$ eikonal directions but in our case we have $4$ eikonal directions $n_1$ to $n_4$, which complicates the color structure in $ |{\cal M}|^2_{eik.}$ quite a lot. The LO soft function is $S_G^{(0)}(\omega) = \delta\left(\omega\right)$. At NLO, we only have 1-emission and the matrix element is given by (see ~\cite{Catani2000} for all the notations,
especially the color factor $T_i$):
\bea
|{\cal M}|^2_{eik.} = -g_s^2  \sum_{i,j = 1, i\ne j}^4 T_i \cdot T_j \frac{n_i\cdot n_j}{n_i \cdot k n_j \cdot k}
\eea
and therefore:
\bea
S_G^{(1)}(\omega) = -g_s^2  \sum_{i,j = 1, i\ne j}^4 T_i \cdot T_j 
 \int  \frac{\mathrm{d}^d k_i}{(2\pi)^{d-1}} \delta(k^2) \theta(k^0) 
\frac{n_i\cdot n_j}{n_i \cdot k n_j \cdot k}
\delta(\omega - \sum_i n_4 \cdot k ) \,.
\eea
We note that if $i = 4$ or $j = 4$, the integration will be scaleless after we use the $\delta$ function
to remove the integral over $n_4 \cdot k$ therefore vanishes. For $i\ne 4$ and $j\ne 4$, following 
~\cite{Becher2012}, we get:
\bea
S_G^{(1)}(\omega) = \sum_{i,j \ne 4,i\ne j } T_i \cdot T_j 
\frac{g_s^2}{(2\pi)^{d-1} } \frac{2\pi^{1-\epsilon}}{\Gamma(1-\epsilon)} \frac{1}{\epsilon}
\left(
\frac{n_{i4} n_{j4}}{2n_{ij}}
\right)^{\epsilon} \,
\frac{1}{\mu} \left( \frac{\omega}{\mu} \right)^{-1-2\epsilon} \,,
\eea
and here we have defined $n_{ij} = n_i \cdot n_j$.

The NNLO contribution contains the double-real emission of $q\bar{q}$ and $gg$, real-virtual  contribution, and exponentiation of the NLO. We consider each of these separately and summarize the results. First consider the double-real emission. In this case we have:
\bea
S_{G,r}^{(2)}(\omega) =  \int \prod_{i=1}^2 \frac{\mathrm{d}^d k_i}{(2\pi)^{d-1}} \delta(k_i^2) \theta(k_i^0) 
\, |{\cal M}|^2_{eik.}[\{k_i\}] \,
\delta(\omega - n_4 \cdot (k_1 + k_2) ) \,,
\eea
The matrix element for $q{\bar q}$ pair is given by Eq.~(95) to Eq.~(97) in~\cite{Catani2000}
and for gluon pair the matrix element is given by Eq.~(108) to Eq.(110) in~\cite{Catani2000}. As a first step we calculate the ${\cal I}_{ij}$ and ${\cal S}_{ij}$ defined in~\cite{Catani2000} by decomposing them to the functions in~\cite{Becher2012}.  For the $q{\bar q}$ pair we have one term that comes from:
\bea
S_{G,r}^{(2)}(\omega) &=&  
 (4\pi \mu^{2\epsilon} \alpha_s)^2 \, T_R \sum_{i,j=1}^4 \, 
\int \prod_{i=1}^2 \frac{\mathrm{d}^d k_i}{(2\pi)^{d-1}} \delta(k_i^2) \theta(k_i^0) 
\,{\cal I}_{ij}(k_1,k_2) 
\delta(\omega - n_4 \cdot (k_1 + k_2) ) \, \nn \\
 &=& 
 (4\pi \mu^{2\epsilon} \alpha_s)^2 \, T_R \sum_{i,j=1}^4 \, 
\int \frac{\mathrm{d}^d q}{(2\pi)^d}\,  \theta(q^0) \theta(q^2)  \, \delta(\omega - n_4 \cdot q ) \, 
\,\tilde{\cal I}_{ij}(q)  \nn \\
&& \times \int \prod_{i=1}^2 \frac{\mathrm{d}^d k_i}{(2\pi)^{d-1}} \delta(k_i^2) \theta(k_i^0) 
\, 
\, 
(2\pi)^{d} \delta^{(d)}( q - k_1 -k_2) 
\nn \\
\eea
Here 
\bea
\,\tilde{\cal I}_{ij}(q) =  - \frac{2 n_{ij}    }{  n_i \cdot q n_j \cdot q \, ( q^2 )}
\eea
where we have neglected the asymmetric part in $k_1$ and $k_2$ which will also contribute. The  third line is nothing but the $2$-body phase space volume for a particle with momentum $q$ decays into $k_1$ and $k_2$, which is (see~\cite{FourParticle}, Eq.~(A.1) and Eq.~(A.9)):
\bea
P_2 = 2^{-3+2\epsilon} \pi^{-1+\epsilon} \frac{\Gamma(1-\epsilon)}{\Gamma(2-2\epsilon)} (q^2)^{-\epsilon}
\eea
Therefore we have:
\bea
S_{G,r}^{(2)}(\omega) 
 &=&  - 
 (4\pi \mu^{2\epsilon} \alpha_s)^2 \, T_R 2^{-2+2\epsilon} \pi^{-1+\epsilon} \frac{\Gamma(1-\epsilon)}{\Gamma(2-2\epsilon)}    
 \frac{1}{(2\pi)^d}
 \nn \\
&& \times \sum_{i \ne j=1}^4 \, 
\int \mathrm{d}^d q\,  \theta(q^0) \, \theta(q^2)  \, \delta(\omega - n_4 \cdot q ) \, 
\,  \frac{ n_{ij}    }{  n_i \cdot q n_j \cdot q \, ( q^2 )^{1+\epsilon} } 
\, 
\, 
\nn \\
\eea
where the second line is $I_2$ of Eq.~(41) in ~\cite{Becher2012}. With regards to the real-virtual contribution, we have one real emission and one loop  which leads to:
\bea
S_{G,v}^{(2)}(\omega) =  \int \frac{\mathrm{d}^d k_1}{(2\pi)^{d-1}} \delta(k_1^2) \theta(k_1^0) 
\, |{\cal M}|^2_{eik.,v}[k_1 ] \,
\delta(\omega - n_4 \cdot k_1  ) \,,
\eea
where the matrix element can be found in Eq.~(26) of ~\cite{Catani2000_1}. Once we figure out the matrix element, the integration is straightforward as NLO. For the soft quark pair emission, the matrix element is given by ~\cite{Catani2000}:
\bea
(4\pi\alpha_s \mu^{2\epsilon})^2 T_R \sum_{i,j=1}^n {\cal I}_{ij} |{\cal M}_{ij}|^2
\eea 
where
\bea
{\cal I}_{ij} =  - \frac{2n_{ij}  k_1 \cdot k_2 +  n_i \cdot (k_1 - k_2)   n_j \cdot (k_1 - k_2)}
{2 (k_1 \cdot k_2)^2  n_i \cdot (k_1+k_2) \,  n_j \cdot (k_1+k_2)  }
\eea

The non-abelian piece of the double gluon matrix element is given by ~\cite{Catani2000}:
\bea
- C_A \, (4\pi\alpha_s \mu^{2\epsilon})^2 \sum_{i,j=1}^n \,  {T_i \cdot T_j} \,  {\cal S}_{ij} |{\cal M}_{ij}|^2
\eea
where ${\cal S}_{ij}$ can be written as 
\bea
{\cal S}_{ij} 
& = & {\cal S}_{ij}^{s.o.} -  \frac{1}{2}\, \frac{n_i \cdot k_1 n_j \cdot k_2 
+ n_j \cdot k_1 n_i \cdot k_2 }{   n_i \cdot (k_1+k_2) \,  n_j \cdot (k_1+k_2)   }  {\cal S}_{ij}^{s.o.}  \nn \\
&&+ (1-\epsilon) {\cal I}_{ij} - (1+\epsilon) \frac{ n_{ij}  }{  k_1 \cdot k_2  n_i\cdot (k_1+k_2) \, n_j \cdot (k_1+k_2) }
\eea
with the strong-ordering limit being:
\bea
{\cal S}_{ij}^{s.o.}  = \frac{n_{ij}}{ k_1 \cdot k_2} \, 
\left(
\frac{1}{n_i \cdot k_1 n_j \cdot k_2} + i \leftrightarrow j
\right)
- \frac{n_{ij}^2}{ n_i \cdot k_1 n_j \cdot k_1 \, n_i \cdot k_2 \, n_j \cdot k_2  }
\eea

With regards to the double gluon emission, first we focus on the term:
\bea
{\cal S}_{ij}^{s.o.}  -  \frac{1}{2}\, \frac{n_i \cdot k_1 n_j \cdot k_2 
+ n_j \cdot k_1 n_i \cdot k_2 }{   n_i \cdot (k_1+k_2) \,  n_j \cdot (k_1+k_2)   }  {\cal S}_{ij}^{s.o.} 
\eea
which can be decomposed to  
\bea
   \,   
 \frac{n_{ij}}{ q^2 } \times \,  
 \left(   
\frac{    
   n_j \cdot (k_1+2k_2)    }
 {   n_j \cdot q \, n_i \cdot k_1 n_j \cdot k_2  } 
  + 
 \frac{    
       n_j \cdot (k_1 -  k_2)   }
 {n_i \cdot q \,  n_j \cdot q \,   n_j \cdot k_2}  \, 
   \right) \, 
 + i \leftrightarrow j
\, 
\eea
and
\bea
- \frac{n_{ij}^2}{ n_i \cdot k_1 n_j \cdot k_1 \, n_i \cdot k_2 \, n_j \cdot k_2  }
\eea
and
\bea
 2\, \frac{n_{ij} }{ 2\,  n_i \cdot q \,  n_j \cdot q  } 
\left(
\frac{   n_{ij}        }{ 2 n_i \cdot k_1  \,  \, n_j \cdot k_2  }
+
  i \leftrightarrow j
\right)
\eea
where we have defined $q = k_1 + k_2$. 

Therefore we have
\bea\boxed{
S_{G, ij, 1}^{gg}(\omega) 
 =  \left( 
 \frac{1}{2\pi}  \right)^{2d-2}
 \Bigg(  2  I_{7,2} + 2 I_6   - I_4 \Bigg)  + 2 I_3^{s.o.}  }
\eea 
where $ I_{7,2}$, $I_6$ and $I_4$ are given in ~\cite{Becher2012} and 
$I_3^{s.o.}$ is given by:
\bea
I_{3}^{s.o.} = 2\, 
\int \mathrm{d}^d q \delta(\omega - n_4 \cdot q) \, 
 \frac{n_{ij} }{ 2\,  n_i \cdot q \,  n_j \cdot q  } 
\, 
\int [ \mathrm{d} k_1 ] [ \mathrm{d} k_2 ] 
\frac{   n_{ij}        }{ 2 n_i \cdot k_1  \,  \, n_j \cdot k_2  }
\delta^{(d)}(q-k_1-k_2) 
\eea
which is:
\bea
&& 2 \pi^{1-\epsilon}
\left[ \frac{1}{2\pi} \right]^{2d-2} \, \frac{\Gamma(-\epsilon)}{ \Gamma(1-2\epsilon) }
\frac{\Omega_{d-3}}{4}
 \frac{1}{\omega}
\left(
\frac{  4  \omega^2 }{n_4^+ n_4^-} 
\right)^{-2\epsilon}  \, 
4^{1+2\epsilon} \,
\sqrt{\pi} \frac{\Gamma\left( \frac{1}{2} - \epsilon  \right)}{ \Gamma(1-\epsilon) }   \,
\frac{1}{2} \sqrt{\pi} \, \frac{\Gamma(-2\epsilon)}{\Gamma\left( \frac{1}{2} - 2\epsilon \right)} 
  \nn \\
&& \times    \int  \, 
   \mathrm{d} u^2  \,  
     \times  
\left(   1-u^2 \right)^{-\epsilon}\, 
\left(    u^2  \right)^{-1-2\epsilon} \,
 {}_2 F_1 \left( -\epsilon,-\epsilon, 1-\epsilon,  1-u^2
\right) \,
 {}_2 F_1 \left( -2\epsilon,-2\epsilon, 1-\epsilon,  u^2
\right)  \nn \\
\eea
Expanding the hypergeometric function and
performing the integration over $u^2$, we get:

\bea
I_3^{s.o.} = \frac{1}{4\pi^2}
\frac{c_R}{16\pi^2}
 \frac{1}{\omega}
\left(
\frac{  4  \omega^2 }{n_4^+ n_4^-} 
\right)^{-2\epsilon}  \, 
\left(
- \frac{1}{2\epsilon^3}
+ \frac{7\pi^2}{12 \epsilon} 
+ \frac{43 }{3} \zeta_3 
+ \frac{121\pi^4}{720} \epsilon
\right)
\eea
Here we have defined:
$
c_R = (4\pi)^{2\epsilon} e^{-2\epsilon \gamma_E}$. Therefore we have:
\bea  
&& 
S_{G, ij ,1}^{gg}(\omega) =  \frac{1}{4\pi^2}
\frac{c_R}{16\pi^2}
 \frac{1}{\omega}
\left(
\frac{  4  \omega^2 }{n_4^+ n_4^-} 
\right)^{-2\epsilon}    \nn \\ 
&&   \times \left[
- \frac{2}{\epsilon^3}
- \frac{2}{\epsilon^2} 
+ \left( -4 + \frac{4}{3} \pi^2  \right) \frac{1}{\epsilon}
+\left(
-8 + \pi^2 + \frac{202}{3}\zeta_3
\right) 
+\left(
-16+2\pi^2+ \frac{124}{3}\zeta_3 + \frac{229}{180}\pi^4 
\right)\epsilon
\right]   \nn \\
\eea
Now we turn to the second line of ${\cal S}_{ij} $ which is:
\bea
 (1-\epsilon) {\cal I}_{ij} + (1+\epsilon) \frac{- 2 n_{ij}  }{  q^2  n_i\cdot q \, n_j \cdot q }
\eea
The second term is given by:
\bea
   (1+\epsilon) \times \left(  - \frac{2 n_{ij}}{  q^2 n_i \cdot q n_j \cdot q   } \right)
    =  (1+\epsilon) \times \tilde{{\cal I}}_{ij}(q) 
\eea
Therefore we find:
\bea
S_{G, ij , 2, +}^{gg}(\omega) = -  (1+\epsilon) \times 2^{-2+2\epsilon} \pi^{-1+\epsilon } \frac{\Gamma(1-\epsilon)}{\Gamma(2-2\epsilon)}
\frac{1}{(2\pi)^d} \times I_2 \,.
\eea
which is:
\bea
  S_{G, ij , 2, +}^{gg}(\omega) = \frac{1}{4\pi^2}
\frac{c_R}{16\pi^2}
 \frac{1}{\omega}
\left(
\frac{  4  \omega^2 }{n_4^+ n_4^-} 
\right)^{-2\epsilon}       \left(
- \frac{1}{\epsilon^2}
- \frac{3}{\epsilon} 
-6 + \frac{\pi^2}{2}
+\left(
-12 + \frac{3\pi^2}{2} + \frac{62}{3}\zeta_3
\right) \epsilon
\right)
\eea
For the first term it can be related to the $|{\cal M}|^2$ for producing a $q{\bar q}$ pair:
\bea
&&\int \frac{\mathrm{d}^d q}{(2\pi)^d} \delta(\omega - n_4 \cdot q)
\int \mathrm{d}{\rm PS} \, {\cal I}_{ij} (2\pi)^d \delta^{(d)}(q - k_1 -k_2) \nn \\
 &=&  2 \int \frac{\mathrm{d}^d q }{(2\pi)^d}
\, \delta(\omega - n_4 \cdot q)
\,{\rm Im} \Pi_{\mu\nu}(q \to q) 
\frac{n_i^\mu}{n\cdot q} \frac{n_j^\nu}{n_j \cdot q} \frac{1}{(q^2)^2}
\eea
The  sign on the right hand side of the equation is determined from the fact that 
the matrix element involving the virtual quark loop has an additional factor of $(-i)^{2} i^2 = 1$ (vertices plus virtual photon propagators). This gives:
\bea
&& 2 
(-8) \frac{1}{(4\pi)^{2-\epsilon}} \Gamma(\epsilon) \
\frac{\Gamma^2(2-\epsilon)}{\Gamma(4-2\epsilon)} \
{\rm Im}[ e^{i \, \pi \epsilon } ]\
\int \frac{ \mathrm{d}^d q }{(2\pi)^d} 
\delta(\omega - n_4 \cdot q)
(q^2 g_{\mu\nu} - q_\mu q_\nu) \
(q^2)^{-\epsilon}\
\frac{n_i^\mu}{n\cdot q} \frac{n_j^\nu}{n_j \cdot q} \frac{1}{(q^2)^2} \ \nn \\
&=&
 2 
(-8) \frac{1}{(4\pi)^{2-\epsilon}} \Gamma(\epsilon) \,
\frac{\Gamma^2(2-\epsilon)}{\Gamma(4-2\epsilon)} \,
{\rm Im}[ e^{i \, \pi \epsilon } ]\,
\int \frac{ \mathrm{d}^d q }{(2\pi)^d} 
\delta(\omega - n_4 \cdot q)
 \,
\frac{n_{ij}}{n\cdot q} \frac{1}{n_j \cdot q} \frac{1}{(q^2)^{1+\epsilon}} \, \nn \\
&=&
 2 
(-8) \frac{1}{(4\pi)^{2-\epsilon}} \Gamma(\epsilon) \,
\frac{\Gamma^2(2-\epsilon)}{\Gamma(4-2\epsilon)} \,
\sin[ { \, \pi \epsilon } ]\,
\frac{1}{(2\pi)^d} \times 
I_2 \, \nn \\
\eea
Therefore we have from the first $(1-\epsilon){\cal I}_{ij} $ term:
\bea
S_{G, ij, 2, qq}^{gg}(\omega) = 
\frac{1}{4\pi^2}
\frac{c_R}{16\pi^2}
 \frac{1}{\omega}
\left(
\frac{  4  \omega^2 }{n_4^+ n_4^-} 
\right)^{-2\epsilon}      \times \\
\times
\left(
-\frac{2}{3 \epsilon^2} 
- \frac{4}{9\epsilon} \, 
- \frac{26}{27}
+ \frac{\pi^2}{3} 
+
\left[ \frac{2\pi^2}{9} 
- \frac{160}{81} \,
+ \frac{124}{9}\zeta_3
\right]\epsilon
\right) \,
\eea
Regarding the $q{\bar q}$ pair emission, we have from ${\cal I}_{ij}$ contribution:
\bea
S_{G, ij}^{q{\bar q}}(\omega) = 
\frac{1}{4\pi^2}
\frac{c_R}{16\pi^2}
 \frac{1}{\omega}
\left(
\frac{  4  \omega^2 }{n_4^+ n_4^-} 
\right)^{-2\epsilon}      \times
\\
\times
\left(
-\frac{2}{3 \epsilon^2} 
- \frac{10}{9\epsilon} \, 
- \frac{56}{27}
+ \frac{\pi^2}{3} 
+
\left[ \frac{5\pi^2}{9} 
- \frac{328}{81} \,
+ \frac{124}{9}\zeta_3
\right]\epsilon
\right) \,
\eea
The real virtual matrix element is given by:
\bea
&& -(g_s \mu^\epsilon)^4 \times\left[ 
-\frac{1}{4\pi^2} \frac{(4\pi)^\epsilon}{\epsilon^2}\,
\frac{\Gamma^3(1-\epsilon)\Gamma^2(1+\epsilon)}{\Gamma(1-2\epsilon)}
\left(
S_2 + S_3
\right) \right] \nn \\
&=&
(4\pi\alpha_s \mu^{2\epsilon})^2 \times\left[ 
\frac{1}{4\pi^2} \frac{(4\pi)^\epsilon}{\epsilon^2}\,
\frac{\Gamma^3(1-\epsilon)\Gamma^2(1+\epsilon)}{\Gamma(1-2\epsilon)}
\left(
S_2 + S_3
\right) \right] 
\eea
where
\bea
S_2 = C_A \cos(\pi \epsilon) \sum_{i \ne j} \left[
S_{ij}(q)
\right]^{1+\epsilon} |{\cal M}_{i,j} (p)  |^2
\eea
and 
\bea
S_3 = 2\sin(\pi \epsilon) \sum_{i\ne j\ne k} S_{ki}(q) 
\left[
S_{ij}(q)
\right]^\epsilon \left( \lambda_{ij} - \lambda_{iq} - \lambda_{jq} \right)
| {\cal M}_{k,i,j}  (p) |^2
\eea
Here 
\bea
&&S_{ij}(q) \equiv  \frac{n_{ij}}{2 n_i \cdot q \, n_j \cdot q}  \\
&& | {\cal M}_{k,i,j}  (p) |^2
= f_{abc} \langle {\cal M}^{0}(p) | T_k^a T_i^b T_j^c | {\cal M}^0 (p) \rangle
\eea
and $\lambda_{AB} = 1$ if $A$ and $B$ are both incoming or outgoing otherwise $\lambda_{AB} = 0$. Integrating over the phase space gives:
\bea
\int [\mathrm{d} q] \delta(\omega - n_4 \cdot q) S_2 
&= & C_A \cos(\pi \epsilon)
\int [\mathrm{d} q] \delta(\omega - n_4 \cdot q)
   \left[
S_{ij}(q) 
\right]^{1+\epsilon} \nn \\
&=&  C_A \cos(\pi \epsilon) \left(\frac{1}{2\pi} \right)^{d-1} \times I_{7,1}
\eea
Therefore its contribution except for the $(4\pi\alpha_s \mu^{2\epsilon})^2$ factor is given by:
\bea
S_{G,rv,2} = 
\frac{C_A}{4\pi^2} 
\frac{c_R}{16\pi^2}
 \frac{1}{\omega}
\left(
\frac{  4  \omega^2 }{n_4^+ n_4^-} 
\right)^{-2\epsilon}      \times
\left(
-\frac{1}{\epsilon^3}
+ \frac{\pi^2}{2 \epsilon} \,
+ \frac{80}{3}\zeta_3 \,
+ \frac{47 \pi^4}{120} \epsilon
\right)
\eea
Now we turn to $S_3$ which only contributes to $4$ or more external hard partons. $S_3$ vanishes since the origin of the $S_3$ term is the $3$-gluon vertex and the anti-symmetric part of the other pieces connected to
$3$ different eikonal lines. Finally, the coupling renormalization term is given by:
\bea
- \frac{\alpha_s}{2\pi} \times \frac{11C_A - 2n_F}{6 \epsilon } \times {\rm NLO}
\eea

\subsection{Summary of results}
We summarize our results as follows. The NLO soft function is given by:
\bea
S_G^{(1)}\left(\omega, \mu\right) = \sum_{i\ne j \,, i,j \ne 4} T_i \cdot T_j  \,
S^{(1)}_{G, ij}\left(\omega, \mu\right)
\eea
where
\bea
S^{(1)}_{G, ij}\left(\omega, \mu\right) =  \frac{\alpha_s}{2\pi} L^{(1)}_{ij} \,
\left(
\frac{2}{\epsilon} - \frac{\pi^2}{6}\epsilon
-\frac{14\zeta_3}{3}\epsilon^2 + O\left(\epsilon^3\right)\right)
\eea
with 
\bea
L^{(1)}_{ij} =   \left(
\frac{n_{i4} n_{j4}}{2n_{ij}}\, 
\right)^{\epsilon}\,
\frac{1}{\mu}
\, \left(
 \frac{\omega \, }{\mu}
\right)^{-1-2\epsilon}
\eea
\subsubsection{NNLO contribution}
The NNLO soft function is given by:
\bea
S_G^{(2)}\left(\omega, \mu\right) =  \frac{1}{4}
\sum_{i\ne j \,, i,j \ne 4 }
\sum_{k\ne l\,, k,l\ne4}
\left\{
T_i\cdot T_j \,, T_k \cdot T_l
\right\}
S_{ij,kl}^{ab.}\left(\omega, \mu\right) \, 
+ \\ + 
\sum_{i\ne j \,, i,j \ne 4} T_i \cdot T_j \left(
S^{C_A}_{ij}\left(\omega, \mu\right) + S^{N_F}_{ij}\left(\omega, \mu\right)  +S^{ren.}_{ij}\left(\omega, \mu\right) 
\right) \,
\eea
The individual contributions are given by:
\bea
S_{ij,kl}^{ab.}\left(\omega, \mu\right) = \left(\frac{\alpha_s}{2\pi}\right)^2\, L_{ij,kl}\,
\left[ 
-\frac{4}{\epsilon^3}
+ \frac{10\pi^2}{3\epsilon} \,
+ \frac{248\zeta_3}{3}
+ \frac{25\pi^4}{18}\epsilon + O\left(\epsilon^2\right)
\right] \,
\eea
\bea
S_{ij}^{C_A}\left(\omega, \mu\right) &=& 
 \frac{\alpha_s^2\,C_A }{4\pi^2} L^{(2)}_{ij} \,  \left[
\frac{11}{6 \epsilon^2}
+ \frac{1}{\epsilon} \left(
\frac{67}{18} - \frac{\pi^2}{6}
\right)
+ \frac{202}{27} 
- \frac{11\pi^2}{12}
- 7 \zeta_3\, \right. \nn \\
&& \left. + \left(
\frac{1214}{81}
-\frac{67\pi^2}{36}
- \frac{11\pi^4}{45}
- \frac{341}{9}\zeta_3
\right)\epsilon + O\left(\epsilon^2\right)
\right] \,
\eea
\bea
S_{ij}^{N_F}\left(\omega, \mu\right) = \frac{\alpha_s^2 T_R N_F}{4\pi^2} L^{(2)}_{ij}\,
\left(-
\frac{2}{3\epsilon^2}\,
- \frac{10}{9\epsilon}\,
-\frac{56}{27}+\frac{\pi^2}{3} + 
\, \right. \nn \\
 \left.
+ \left[
\frac{5\pi^2}{9} - \frac{328}{81} + \frac{124}{9}\zeta_3
\right]\epsilon  + O\left(\epsilon^2\right)
\right)
\eea

\bea
S^{ren.}_{ij}\left(\omega, \mu\right) = \frac{\alpha_s^2}{4\pi^2}\frac{11 C_A - 2 N_F}{3}\,
L_{ij}^{(1)}
\left(-\frac{1}{\epsilon^2} + \frac{\pi^2}{12} + \frac{7}{3}\zeta_3 \epsilon + O\left(\epsilon^2\right)
\right)
\eea
Here
\bea
L^{(2)}_{ij} = \frac{1}{\omega} \left(
\frac{4\omega^2}{n_4^+ n_4^- \, } \frac{1}{ \mu^2}
\right)^{-2\epsilon}
=  \left(
\frac{n_{i4} n_{j4}}{2n_{ij}}\, 
\right)^{2\epsilon}\,
\frac{1}{\mu}
\, \left(
 \frac{\omega \, }{\mu}
\right)^{-1-4\epsilon}
\eea
and 
\bea
L_{ij,kl} 
=  \left(
\frac{n_{i4} n_{j4}}{2n_{ij}}\, 
\right)^{\epsilon}\,
\left(
\frac{n_{k4} n_{l4}}{2n_{kl}}\, 
\right)^{\epsilon}\,
\frac{1}{\mu}
\, \left(
 \frac{\omega \, }{\mu}
\right)^{-1-4\epsilon}
\eea

\subsection{Plus expansion of our results in $\epsilon$}
We perform the plus expansion of our results with the use of the 
formula: 
\bea
\frac{1}{x^{1+\alpha}} = -\frac{\delta\left(x\right)}{\alpha}+
\sum_{n=0}^{\infty}\left(-\alpha\right)^n\left(\frac{\theta{\left(x\right)}\ln^n\left(x\right)}{x}\right)_{+}
\eea
We use the conventions:
$\left(ij\right) = \ln\left(\frac{n_{i4}n_{j4}}{2n_{ij}}\right)$, $\left(ijkl\right) = \ln\left(\frac{n_{i4}n_{j4}n_{k4}n_{l4}}{4n_{ij}n_{kl}}\right)$.

\bea
S_G^{(1)}\left(\omega,\mu\right)=
\sum_{i\ne j; i,j \ne 4} (T_i\cdot T_j)\ \frac{\alpha_S}{2\pi}\left[ -\frac{1}{\epsilon^2}\delta\left(\omega\right)+\frac{1}{\epsilon}\left(2\left(\frac{\theta\left(\omega\right)}{\omega}\right)_+-\delta\left(\omega\right)\left(ij\right)\right) + \, \right. \nn \\
\left.  +
\delta\left(\omega\right)\left(\frac{\pi^2}{12}-
\frac{1}{2}\left(ij\right)^2 \right) + 2\left(ij\right)\left(\frac{\theta\left(\omega\right)}{\omega}\right)_+-
4\left(\frac{\ln\left(\frac{\omega}{\mu}\right)}{\omega}\right)_+\right]
\eea

The divergent and finite parts are given by:
\bea 
S^{ab}_{ij, kl}|_{Divergent}= 
\frac{\alpha^2_S}{4\pi^2}\sum_{i\ne j,  i,j \ne 4}^4\sum_{k\ne l,  k,l \ne 4}^4  \{ {T}_i\cdot {T}_j, {T}_k\cdot {T}_l\}\left[\frac{\delta(\omega)}{2\epsilon^4} + \frac{1}{2\epsilon^3}\left( \delta(\omega)(ijkl)-4\left(\frac{\theta(\omega)}{\omega}\right)_+\right)+ \right. \nn \\
\left.  \   +
\frac{1}{\epsilon^2}\left( -2\left(\frac{\theta(\omega)}{\omega}\right)_+(ijkl)+\delta(\omega)\left(-\frac{5\pi^2}{12} + \frac{1}{4}(ijkl)^2\right)+8\left(\frac{\ln\left(\frac{\omega}{\mu}\right)}{\omega}\right)_+\right)+\right. \nn \\
\left.  \   
+ \frac{1}{\epsilon}\left(\delta(\omega)\left(-\frac{31\zeta_3}{3}-\frac{5\pi^2(ijkl)}{12}+\frac{(ijkl)^3}{12}\right)+\left(\frac{\theta(\omega)}{\omega}\right)_+\left(\frac{5\pi^2}{3} - (ijkl)^2\right)\right)+
\right. \nn \\
\left.  \  +
\frac{1}{\epsilon}\left(8(ijkl)\left(\frac{\ln\left(\frac{\omega}{\mu}\right)}{\omega}\right)_+  -16\left(\frac{\ln^2\left(\frac{\omega}{\mu}\right)}{\omega}\right)_+\right)\right] \ \ \ \ \ \ \ 
\eea
\bea 
S^{ab}_{ij, kl}|_{Finite}= 
\frac{\alpha^2_S}{4\pi^2}\sum_{i\ne j,  i,j \ne 4}^4\sum_{k\ne l,  k,l \ne 4}^4  \{ {T}_i\cdot {T}_j, {T}_k\cdot {T}_l\}\left[\delta(\omega)\left(
\frac{(ijkl)^4}{48}-\frac{25\pi^4}{144}-\frac{31\zeta_3}{3}(ijkl)-
\frac{5\pi^2}{24}(ijkl)^2\right)+  \right. \nn \\
\left.  \  
+ \left(\frac{\theta(\omega)}{\omega}\right)_+\left(\frac{124\zeta_3}{3}+\frac{5\pi^2}{3}(ijkl)-\frac{(ijkl)^3}{3}\right)+
\left(\frac{\ln\left(\frac{\omega}{\mu}\right)}{\omega}\right)_+
\left(4(ijkl)^2-\frac{20\pi^2}{3}\right) -  \right. \nn \\
\left.  \  
 - 16 (ijkl)\left(\frac{\ln^2\left(\frac{\omega}{\mu}\right)}{\omega}\right)_+ 
+ \frac{64}{3}\left(\frac{\ln^3\left(\frac{\omega}{\mu}\right)}{\omega}\right)_+\right] \ \ \ \ \ \ \ \
\eea

\bea
S^{C_A}_{ij}|_{Divergent} = \frac{\alpha_S^2C_A}{4\pi^2}\sum_{i\ne j;  i,j \ne 4}^4 \left({T}_i\cdot {T}_j\right) \left[-\frac{11\delta(\omega)}{24\epsilon^3} +\frac{\delta(\omega)}{\epsilon^2}\left(\frac{\pi^2}{24}-\frac{67}{72}\right)+ \right. \nn \\
\left. \ 
+ \frac{\delta(\omega)}{\epsilon}\left(\frac{11\pi^2}{48}+\frac{7\zeta_3}{4}-\frac{101}{54} + \left(\frac{\pi^2}{12}-\frac{67}{36}\right)\cdot(ij)-\frac{11}{12}(ij)^2\right)+ \right. \nn \\
\left.  \  
+ \frac{1}{\epsilon}
\left(\frac{67}{18}-\frac{\pi^2}{6} + \frac{11}{3}(ij)\right)\left(\frac{\theta(\omega)}{\omega}\right)_+ - \frac{22}{3\epsilon}\left(\frac{\ln\left(\frac{\omega}{\mu}\right)}{\omega}\right)_+\right]\ \ \ \ \ \
\eea
\bea 
S^{C_A}_{ij}|_{Finite}=\frac{\alpha_S^2 C_A}{4\pi^2}\sum_{i\ne j;  i,j \ne 4}^4 \left({T}_i\cdot {T}_j\right) \left[\left(\frac{\theta(\omega)}{\omega}\right)_+\left( \frac{202}{27}-\frac{11\pi^2}{12}-7\zeta_3+ (ij)(\frac{67}{9}-\frac{\pi^2}{3})+\frac{11}{3}(ij)^2\right) + \right. \nn \\
\left.  \  
+
 \delta(\omega)\left(\frac{341\zeta_3}{36} + 
\frac{11\pi^4}{180}+
\frac{67\pi^2}{144}-\frac{607}{162} +
(ij)\left(\frac{7\zeta_3}{2}+\frac{11\pi^2}{24}-\frac{101}{27}\right)+(ij)^2\left(\frac{\pi^2}{12}-\frac{67}{36}\right) 
- \frac{11}{18}(ij)^3 \right) +  \right. \nn\\
\left.\  
+ \left(\frac{\ln\left(\frac{\omega}{\mu}\right)}{\omega}\right)_+\left(\frac{2\pi^2}{3}-\frac{134}{9} -\frac{44}{3}(ij)\right)+\frac{44}{3}\left(\frac{\ln^2(\frac{\omega}{\mu})}{\omega}\right)_+\right] \ \ \ \ \ \ 
\eea
\bea 
S^{N_F}_{ij}|_{Divergent} =
\frac{\alpha^2_S T_R N_F}{4\pi^2}\sum_{i\ne j;  i,j \ne 4}^4 \left({T}_i\cdot {T}_j\right) \left[\frac{\delta(\omega)}{6\epsilon^3}+
\frac{1}{\epsilon^2}\left(\left[ \frac{5}{18} + \frac{(ij)}{3} \right]\delta(\omega) - \frac{2}{3}\left(\frac{\theta(\omega)}{\omega}\right)_+\right) +\right. \nn\\
\left.\   +
 \frac{1}{\epsilon}\left(\left(\frac{14}{27}-\frac{\pi^2}{12}+\frac{5}{9}(ij)+\frac{(ij)^2}{3}\right) \delta(\omega)-\left(\frac{10}{9}+\frac{4(ij)}{9}\right)\left(\frac{\theta(\omega)}{\omega}\right)_++\frac{8}{3}\left(\frac{\ln\left(\frac{\omega}{\mu}\right)}{\omega}\right)_+\right)\right]\ \ \ \ \ \
\eea
\bea 
S^{N_F}_{ij}|_{Finite} =
\frac{\alpha^2_S T_R N_F}{4\pi^2}\sum_{i\ne j;  i,j \ne 4}^4 \left({T}_i\cdot {T}_j\right) \left[ 
\left(\frac{\theta(\omega)}{\omega}\right)_+\left(\left(\frac{\pi^2}{3}-\frac{56}{27}\right)-\frac{20}{9}(ij)-\frac{4}{3}(ij)^2\right) +\right. \nn\\
\left.\ 
+ \delta(\omega)\left(\left(\frac{82}{81}-\frac{5\pi^2}{36}-\frac{31\zeta_3}{9}\right)   +(ij)\left(\frac{28}{27}-\frac{\pi^2}{6}\right)+\frac{5}{9}(ij)^2+\frac{2}{9}(ij)^3\right) +\right. \nn\\
\left.\ + \left(\frac{\ln\left(\frac{\omega}{\mu}\right)}{\omega}\right)_+\left(\frac{40}{9}+\frac{16}{3}(ij)\right)-\frac{16}{3}\left(\frac{\ln^2\left(\frac{\omega}{\mu}\right)}{\omega}\right)_+\right]
\eea
\bea 
S^{ren}_{ij}|_{Divergent} = \frac{\alpha^2_S}{4\pi^2}\cdot \frac{11C_A-2N_F}{3}\sum_{i\ne j;  i,j \ne 4}^4 \left({T}_i\cdot {T}_j\right) \left[\frac{\delta(\omega)}{2\epsilon^3} + 
\frac{1}{\epsilon^2}\left(\frac{(ij)\delta(\omega)}{2} - \left(\frac{\theta(\omega)}{\omega}\right)_+\right)+\right. \nn\\
\left.\
+ \frac{1}{\epsilon}\left(\left(
\frac{(ij)^2}{4}-\frac{\pi^2}{24} \right)\delta(\omega)+2\left(\frac{\ln\left(\frac{\omega}{\mu}\right)}{\omega}\right)_+-(ij) \left(\frac{\theta(\omega)}{\omega}\right)_+\right)\right] \ \ \ 
\eea
\bea 
S^{ren}_{ij}|_{Finite} = \frac{\alpha^2_S}{4\pi^2}\cdot \frac{11C_A-2N_F}{3}\sum_{i\ne j;  i,j \ne 4}^4 \left({T}_i\cdot {T}_j\right) \cdot \left[\left(\frac{\pi^2}{12}-\frac{(ij)^2}{2}\right)\left(\frac{\theta(\omega)}{\omega}\right)_++
\right. \nn\\
\left.\
+
\delta(\omega)\left(-\frac{7\zeta_3}{6}-\frac{\pi^2(ij)}{24} + \frac{(ij)^3}{12}\right) + 2(ij)
\left(\frac{\ln\left(\frac{\omega}{\mu}\right)}{\omega}\right)_+
-2\left(\frac{\ln^2\left(\frac{\omega}{\mu}\right)}{\omega}\right)_+\right]
\eea

\section{Renormalization and Resummation}\label{sec:Renorm_Resum}

In this Section we perform the renormalization and resummation which can be conveniently done in
Laplace space. Therefore, we perform the Laplace transformation first:
\bea
\tilde{S}\left(L, \mu\right) = \int^\infty_0 \mathrm{d}\omega \ e^{-\rho\omega} S\left(\omega, \mu\right), \ \rho = \frac{1}{\xi e^{\gamma_E}}
\eea
The Laplace-transformed soft function is expressed in terms of the logarithm of the Laplace-space variable $L = \ln\left(\frac{\mu}{\xi}\right)$. The Laplace transformation is performed using an identity: 
\bea
\int^{\infty}_0 \mathrm{d}\omega e^{-\rho\omega}\omega^{-1-n\epsilon} = e^{-n\epsilon\gamma_E}\Gamma(-n\epsilon)\xi^{-n\epsilon}
\eea
with $\gamma_E\approx0.577$ being the Euler constant. For the further convenience, we split the scale $L$ and kinematics parts $l_{ij}$ of the logarithms:
\bea
L_{ij} = \ln\left(\frac{\mu}{\xi}\sqrt{\frac{n_{i4}n_{j4}}{2 n_{ij}}}\right) = \underbrace{\ln\left(\frac{\mu}{\xi}\right)}_{= L} + \underbrace{\ln\left(\sqrt{\frac{n_{i4}n_{j4}}{2 n_{ij}}}\right)}_{= l_{ij}}  = L + l_{ij}
\eea
\bea
L_{ij ,kl}  = \ln\left(\frac{\mu}{\xi}\sqrt[4]{\frac{n_{i4}n_{j4}n_{k4}n_{l4}}{4 n_{ij}n_{kl}}}\right)  = \underbrace{\ln\left(\frac{\mu}{\xi}\right)}_{= L} + \underbrace{\frac{1}{2}\ln\left(\sqrt{\frac{n_{i4}n_{j4}}{2 n_{ij}}}\right)}_{= \frac{1}{2}l_{ij}} + \underbrace{\frac{1}{2}\ln\left(\sqrt{\frac{n_{k4}n_{l4}}{2 n_{kl}}}\right)}_{= \frac{1}{2}l_{kl}}  =  \\ =   L + \frac{1}{2}l_{ij} + \frac{1}{2}l_{kl} 
\eea

NLO contribution takes the form:
\bea
\tilde{S}_G^{(1)}\left(\xi, \mu\right) = \frac{\alpha_S}{2\pi}\sum_{i\ne j;  i,j \ne 4}^4 \left({T}_i\cdot {T}_j\right) 
\tilde{S}_{G,ij}^{(1)}\left(\xi, \mu\right) \ \ \ \ 
\eea

\bea
\tilde{S}_{G,ij}^{(1)}\left(\xi, \mu\right) = 
  - \frac{1}{\epsilon^2} - \frac{2L_{ij}}{\epsilon} - 2 L^2_{ij} - \frac{\pi^2}{4} + \epsilon\left(\frac{4L^2_{ij}}{3} - \frac{\pi^2 L_{ij}}{2} - \frac{\zeta_3}{3}\right) - \\
-\epsilon^2\left(\frac{2 L^4_{ij}}{3} + \frac{\pi^2}{2} L^2_{ij} + \frac{2\zeta_3}{3} L_{ij} + \frac{19\pi^4}{480}\right) 
+O(\epsilon^3) 
\eea

NNLO contribution consists of four parts:
\bea
\tilde{S}_G^{(2)}\left(\xi, \mu\right) = \tilde{S}^{ab.}\left(\xi, \mu\right) + \tilde{S}^{C_A}\left(\xi, \mu\right) + 
\tilde{S}^{N_f}\left(\xi, \mu\right) +
\tilde{S}^{Coup.Ren.}\left(\xi, \mu\right)
\eea
The bare NNLO soft function has a form:
\bea\label{CompleteTogether}
\tilde{S}_G\left(\xi\right)^{Bare} = \hat{1} + \tilde{S}_G^{(1)}\left(\xi, \mu\right) + \tilde{S}_G^{(2)}\left(\xi, \mu\right) + O\left(\alpha_S^3\right)
\eea
here $\hat{1}$ is a unit matrix in a color space. The complete NNLO soft function is obtained by multiplying an Eqn.\ref{CompleteTogether} by the tree-level color matrix. These matrices for all channel are provided in Section~\ref{Color}. The abelian part of NNLO has a form:
\bea
\tilde{S}^{ab.}\left(\xi, \mu\right) = \tilde{S}^{ab.}\left(\xi, \mu\right)_{Divergent} + \tilde{S}^{ab.}\left(\xi, \mu\right)_{Finite}
\eea
\bea
\tilde{S}^{ab.}\left(\xi, \mu\right)_{Divergent} = \frac{\alpha_S^2 }{4\pi^2}\sum_{i\ne j,  i,j \ne 4}^4\sum_{k\ne l,  k,l \ne 4}^4  \{ {T}_i\cdot {T}_j, {T}_k\cdot {T}_l\}\times \\
\times\left[ \frac{1}{4\epsilon^4} +  \frac{L_{ij, kl}}{\epsilon^3} +    \frac{2L^2_{ij, kl}}{\epsilon^2} + \frac{\pi^2}{8\epsilon^2} +
\frac{8L^3_{ij, kl}}{3\epsilon} 
+ \frac{\pi^2 L_{ij, kl}}{2\epsilon} + 
\frac{\zeta_3}{6\epsilon} \right]
\eea
\bea
\tilde{S}^{ab.}\left(\xi, \mu\right)_{Finite}= \frac{\alpha_S^2 }{4\pi^2}\sum_{i\ne j,  i,j \ne 4}^4
\sum_{k\ne l,  k,l \ne 4}^4 \{ {T}_i\cdot {T}_j, {T}_k\cdot {T}_l\}\times \\ \times\left[\frac{8 L^4_{ij, kl}}{3}+\pi^2 L^2_{ij, kl} + \frac{2\zeta_3 L_{ij, kl}}{3} + 
\frac{17\pi^4}{480}\right]
\eea
The double-gluon emission and real-virtual contribution is given by:
\bea
\tilde{S}^{C_A}\left(\xi, \mu\right) = \tilde{S}^{C_A}\left(\xi, \mu\right)_{Divergent} + \tilde{S}^{C_A}\left(\xi, \mu\right)_{Finite}
\eea
\bea
\tilde{S}^{C_A}\left(\xi, \mu\right)_{Divergent} = \frac{\alpha_S^2 C_A}{4\pi^2}\sum_{i\ne j;  i,j \ne 4}^4 \left({T}_i\cdot {T}_j\right)
\left[ -\frac{11}{24\epsilon^3} - 
\frac{11L_{ij}}{6\epsilon^2}-\frac{67}{72\epsilon^2} + 
\, \right. \nn \\
\left.
+ \frac{\pi^2}{24\epsilon^2}-\frac{11L^2_{ij}}{3\epsilon}-\frac{67 L_{ij}}{18\epsilon} + \frac{\pi^2 L_{ij}}{6 \epsilon}  + \frac{7\zeta_3}{4\epsilon} - \frac{101}{54\epsilon} - \frac{55\pi^2}{144\epsilon} \right]
\eea
\bea
\tilde{S}^{C_A}\left(\xi, \mu\right)_{Finite} = \frac{\alpha_S^2 C_A}{4\pi^2}\sum_{i\ne j;  i,j \ne 4}^4 \left( {T}_i\cdot {T}_j \right)
\left[-\frac{44L^3_{ij}}{9} + \frac{\pi^2 L^2_{ij}}{3} -\frac{67 L^2_{ij}}{9}  + 7\zeta_3 L_{ij} - \, \right. \nn \\
\left. - \frac{55\pi^2}{36}L_{ij} - \frac{202}{27}L_{ij} 
- \frac{11\zeta_3}{36}  + \frac{7\pi^4}{60}-\frac{335\pi^2}{432} - \frac{607}{162} \right]
\eea
The $q\bar{q}$ emission contribution has a form:
\bea
\tilde{S}^{N_F}\left(\xi, \mu\right) = \tilde{S}^{N_F}\left(\xi, \mu\right)_{Divergent}  + \tilde{S}^{N_F}\left(\xi, \mu\right)_{Finite}
\eea
\bea
\tilde{S}^{N_F}\left(\xi, \mu\right)_{Divergent} = \frac{\alpha_S^2 N_F T_R}{4\pi^2}\sum_{i\ne j;  i,j \ne 4} \left({T}_i\cdot {T}_j\right)\left[\frac{1}{6\epsilon^3} + \frac{2 L_{ij}}{3\epsilon^2} + \frac{5}{18\epsilon^2}  + \frac{4 L^2_{ij}}{3\epsilon} + \, \right. \nn \\
\left. + \frac{10 L_{ij}}{9\epsilon}+\frac{14}{27\epsilon}
+ \frac{5\pi^2}{36 \epsilon}\right]
\eea
\bea
\tilde{S}^{N_F}\left(\xi, \mu\right)_{Finite} = \frac{\alpha_S^2 N_F T_R}{4\pi^2}\sum_{i\ne j;  i,j \ne 4} \left({T}_i\cdot {T}_j\right)
\left[ \frac{16 L^3_{ij}}{9} + \frac{20 L^2_{ij}}{9} + \frac{5\pi^2 L_{ij}}{9} + \, \right. \nn \\
\left.  + \frac{56 L_{ij}}{27} + \frac{\zeta_3}{9} +\frac{25\pi^2}{108} 
+ \frac{82}{81}\right]
\eea
Finally, the renormalization of the coupling constant is:
\bea
\tilde{S}^{Coup.Ren.}\left(\xi, \mu\right) = \tilde{S}^{Coup.Ren.}\left(\xi, \mu\right)_{Divergent} + \tilde{S}^{Coup.Ren.}\left(\xi, \mu\right)_{Finite}
\eea
\bea
\tilde{S}^{Coup.Ren.}\left(\xi, \mu\right)_{Divergent} = \frac{\alpha_S^2 \beta_0}{4\pi^2}\sum_{i\ne j;  i,j \ne 4}\left({T}_i\cdot {T}_j\right)\left[\frac{1}{2\epsilon^3} + \frac{L_{ij}}{\epsilon^2}+\frac{L^2_{ij}}{\epsilon} +\frac{\pi^2}{8\epsilon}\right]
\eea
\bea
 \tilde{S}^{Coup.Ren.}\left(\xi, \mu\right)_{Finite} = \frac{\alpha_S^2 \beta_0}{4\pi^2}\sum_{i\ne j;  i,j \ne 4}\left({T}_i\cdot {T}_j\right)\left[\frac{2L^3_{ij}}{3} + \frac{\pi^2 L_{ij}}{4} +\frac{\zeta_3}{6}\right]
\eea
The soft function is multiplicatively renormalized as:
\bea
\tilde{S}_G\left(\xi, \mu\right)^{Renormalized} = Z\left(\xi, \mu\right)\tilde{S}_G\left(\xi\right)^{Bare}\left(Z\left(\xi, \mu\right)\right)^\dagger
\eea
\bea 
Z\left(\xi, \mu\right) = 1 - \frac{\alpha_S}{2\pi}\sum_{i\ne j; i,j\ne 4}\left({T}_i\cdot {T}_j\right)\left(\frac{\Gamma_0}{2\epsilon^2} + \frac{\left(2L_{ij} - i\pi\lambda_{ij}\right)\Gamma_0 + 2\gamma_0}{2\epsilon}\right) + 
\eea
\bea
+ \left(\frac{\alpha_S}{2\pi}\right)^2\sum_{i\ne j; i,j\ne 4}\left({T}_i\cdot {T}_j\right)\times
\\
\times\left(\frac{3\beta_0\Gamma_0}{16\epsilon^3} - \frac{\Gamma_1 - \beta_0\left(\left(2L_{ij} - i\pi\lambda_{ij}\right)\Gamma_0 + 2\gamma_0\right)}{8\epsilon^2} 
- \frac{\left(2L_{ij} - i\pi\lambda_{ij}\right)\Gamma_1 + 2\gamma_1}{4\epsilon}\right) +
\eea
\bea
+ \frac{1}{4}\left(\frac{\alpha_S}{2\pi}\right)^2\sum_{i\ne j; i,j\ne 4}\sum_{k\ne l; k,l\ne 4}\{T_i\cdot T_j, T_k\cdot T_l\}\times \\
\times\left(\frac{\Gamma_0}{2\epsilon^2} + \frac{\left(2L_{ij} - i\pi\lambda_{ij}\right)\Gamma_0 + 2\gamma_0}{2\epsilon}\right)\left(\frac{\Gamma_0}{2\epsilon^2} + \frac{\left(2L_{kl} - i\pi\lambda_{kl}\right)\Gamma_0 + 2\gamma_0}{2\epsilon}\right)
\eea
The anomalous dimension is given by:
\bea
\Gamma\left(\xi, \mu\right) = \sum_{i,j = 1; i\ne j}^4 T_i\cdot T_j \left(2L_{ij} - i\lambda_{ij}\right)\Gamma + 2 \gamma
\eea
\bea
\Gamma = \sum_{n=1}^{\infty}\Gamma_n\left(\frac{\alpha_S}{2\pi}\right)^n, \gamma = \sum_{n=1}^{\infty}\gamma_n\left(\frac{\alpha_S}{2\pi}\right)^n
\eea
With the first two coefficients given by:
\bea 
\Gamma_0 = -1, \ \Gamma_1 = \left(\frac{\pi^2}{6} - \frac{67}{18}\right)C_A + \frac{10}{9}n_F T_R
\eea
\bea
\gamma_0 = 0, \ \gamma_1 = C_A\left(\frac{11\pi^2}{144} + \frac{7\zeta_3}{4} - \frac{101}{54}\right) + n_F T_R\left(\frac{14}{27} - \frac{\pi^2}{36}\right)
\eea
Both the finite parts and an interplay of NLO and the $Z_{NLO}$ (the term in $Z\sim\alpha_S$) contribute to the renormalized soft function:
\bea
\tilde{S}_G\left(\xi, \mu\right)^{Renormalized} = 1 + \left(\tilde{S}^{(1)}_G\left(\xi, \mu\right)\right)_{Finite} + \left(\tilde{S}^{(2)}_G\left(\xi, \mu\right)\right)_{Finite} + \\ + \left(\tilde{S}^{(1)}_G\left(\xi, \mu\right)\cdot 
Z_{NLO}^\dagger\right)_{Finite} + \left(
Z_{NLO}\cdot\tilde{S}^{(1)}_G\left(\xi, \mu\right)\right)_{Finite}
\eea
where:
\bea
\left(\tilde{S}^{(1)}_G\left(\xi, \mu\right)\cdot 
Z_{NLO}^\dagger\right)_{Finite} + \left(
Z_{NLO}\cdot\tilde{S}^{(1)}_G\left(\xi, \mu\right)\right)_{Finite} =  \\ = \frac{\alpha_S^2 }{4\pi^2}\sum_{i\ne j,  i,j \ne 4} 
\sum_{k\ne l,  k,l \ne 4} \{ {T}_i\cdot {T}_j, {T}_k\cdot {T}_l\}  \left[-\frac{19\pi^4}{960} -\frac{2\zeta_3 L_{ij}}{3} - \frac{\pi^2 L^2_{ij}}{4} - \frac{\pi^2 L_{ij}L_{kl}}{2} + \, \right. \nn \\
\left. + \frac{4L_{ij}L^2_{kl}}{3}
- \frac{L^4_{ij}}{3}\right]
\eea

The RG equation has the form:
\bea
\frac{\mathrm{d}}{\mathrm{d}  \ln\left(\mu\right)}\tilde{S}_G\left(\xi, \mu\right) = \Gamma\left(\xi, \mu\right)\tilde{S}_G\left(\xi, \mu\right) + \tilde{S}_G\left(\xi, \mu\right)\Gamma\left(\xi, \mu\right)^{\dagger}\label{RG}
\eea

and its solution is given by:
\begin{equation}
\begin{split}
\tilde{S}_G\left(\xi, \mu\right)^{Resummed} = 1 +  
\frac{\alpha_S}{2\pi}\sum_{i,j = 1, i\ne j}^4  T_i\cdot T_j\left[2\Gamma_0L^2 + 4L\left(\Gamma_0l_{ij} +\gamma_0 \right)\right] + \tilde{S}^{(1)}_{G}\left(L=0\right) + \ \ \ \ \ \ \  \\
+
\left(\frac{\alpha_S}{2\pi}\right)^2\sum_{i,j = 1, i\ne j}^4  T_i\cdot T_j\left[\frac{2\beta_0\Gamma_0L^3}{3} + 2L^2\beta_0\left(\Gamma_0l_{ij} +\gamma_0 \right)\right] + L\beta_0 \tilde{S}^{(1)}_{G}\left(L = 0\right) + \ \ \ \ \ \ \ \\
+ \left(\frac{\alpha_S}{2\pi}\right)^2\sum_{i,j = 1, i\ne j}^4  T_i\cdot T_j\left(2\Gamma_1L^2 + 4L\left(l_{ij} +\gamma_1 \right) \right) 
+ \frac{1}{4}\left(\frac{\alpha_S}{2\pi}\right)^2\sum_{i,j = 1, i\ne j}^4\sum_{k,l = 1, k\ne l}^4  \{T_i\cdot T_j, T_k\cdot T_l\}\times \\
\times\left(2\Gamma_0L^2 + 4L\left(\Gamma_0l_{ij} +\gamma_0 \right) + \tilde{S}^{(1)}_{G,ij}\left(L = 0\right)\right)\left(2\Gamma_0L^2 + 4L\left(\Gamma_0l_{kl} +\gamma_0 \right) + \tilde{S}^{(1)}_{G, kl}\left(L = 0\right)\right) - \\
-4\pi\left(\frac{\alpha_S}{2\pi}\right)^2\sum_{i\ne j \ne k \ne 4} f^{abc} T^a_i T^b_j T^c_l
\left(\Gamma_0\lambda_{ij}\left( 2 L^2\Gamma_0 l_{jl}    + L \tilde{S}^{(1)}_{G, jl}\left(L = 0\right)\right)  +  \tilde{S}^{NNLO}_{G, ijl}\left(L = 0\right) \right)  + \tilde{S}_G^{(2)}\left(L = 0\right)
\end{split}
\end{equation}

Note that $\tilde{S}^{NNLO}_{G, ijl}\left(\xi, \mu\right) = 0$ as we discussed in Section~\ref{sec:Derivation}.  In the next Section we provide the numerical results and perform the threshold resummation. 

\section{Numerical Results}\label{sec:Numerical_Results}
In this section we perform a numerical integration of the Eqn.~\ref{RG}.  The solution is given by:

\bea
U\left(\mu; \ln\left(\frac{\mu_s}{\xi}\right)\right) = \sum_{i,j = 1, i\ne j}^4 2T_i \cdot T_j\int^{\alpha_S\left(\mu\right)}_{\alpha_S\left(\mu_s\right)}
\frac{\mathrm{d}\alpha_S}{\beta\left(\alpha_S\right)}\Gamma\left(\alpha_S\right)\int^{\alpha_S}_{\alpha_S\left(\mu_s\right)}\frac{\mathrm{d}\alpha_S^{\prime}}{\beta\left(\alpha_S^{\prime}\right)} + \nn \\ + \int^{\alpha_S\left(\mu\right)}_{\alpha_S\left(\mu_s\right)}
\frac{\mathrm{d}\alpha_S}{\beta\left(\alpha_S\right)} \left(\sum_{i,j = 1, i\ne j}^4 T_i \cdot T_j\left(2l_{ij} -i\lambda_{ij}\right)\Gamma\left(\alpha_S\right)  + 2\gamma\left(\alpha_S\right)\right) + \nn \\ +
\sum_{i,j = 1, i\ne j}^4 2T_i \cdot T_j\ln\left(\frac{\mu_s}{\xi}\right)\int^{\alpha_S\left(\mu\right)}_{\alpha_S\left(\mu_s\right)}
\frac{\mathrm{d}\alpha_S}{\beta\left(\alpha_S\right)}\Gamma\left(\alpha_S\right)
\eea

$$
S^{Resummed}_G = \exp\left(U\left(\mu; \ln\left(\frac{\mu_s}{\xi}\right)\right)
\right)S^{Renormalized}_G\left(\alpha\left(\mu_s\right), \ln\left(\frac{\mu_s}{\xi}\right)\right)\exp\left(U\left(\mu; \ln\left(\frac{\mu_s}{\xi}\right)\right)
\right)^\dagger
$$

We introduce the following reference vectors:
$$
\begin{cases} 
n_1^{\prime}  = \left(1, 0, 0, 1\right)\\ 
n_2^{\prime}  = \left(1, 0, 0, -1\right) \\
n_3^{\prime}  = \left(1, 0, 1, 0\right)  \\
n_4^{\prime}  = \left(1, 0, -1, 0\right) 
\end{cases}
$$
and provide the ratios of the traces of NLO to NNLO soft functions as well as the ratios of NNLO to NLO contributions in FIGs.1-3 as functions of the logarithm $L$ for each of the color channels.

In FIGs.4-6 we provide the corresponding results for the resummed soft function. Starting from the value of $\alpha_S\left(\mu_s\right)$, we perform an evolution to the final value $\alpha_S=0.118$. The matrix exponentiation was performed by the Mathematica function MatrixExp. Our numerical code is publicly available on GitHub \footnote{\href{https://github.com/BalytskyiJaroslaw/NNLLs}{https://github.com/BalytskyiJaroslaw/NNLLs}}{\href{https://github.com/BalytskyiJaroslaw/NNLLs}{repository}}.

\begin{figure}[ht]
\begin{center}
\begin{minipage}[b]{0.47\linewidth}
\includegraphics[width=\textwidth,angle=0]{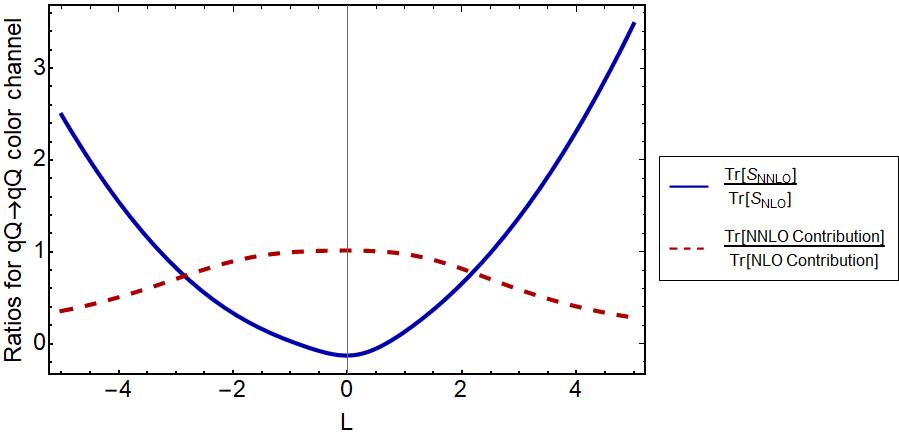}
\end{minipage}
\hspace{0.5cm}
\begin{minipage}[b]{0.47\linewidth}
\includegraphics[width=1.01\textwidth,angle=0]{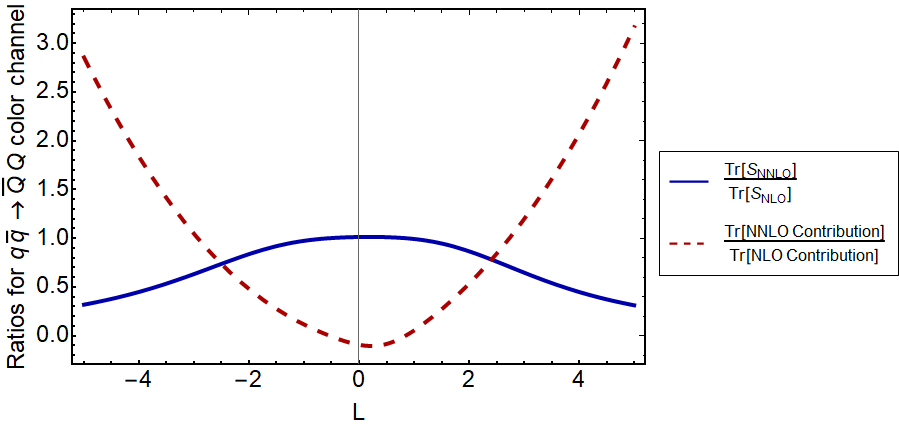}
\end{minipage}
\end{center}
\vspace{-0.5cm}
\caption{The ratios of NNLO to NLO soft function and NNLO to NLO contribution for the $qQ\rightarrow qQ$ and $q\bar{q}\rightarrow\bar{Q}Q$ color channels.} \label{Ren1} 
\end{figure}

\begin{figure}[ht]
\begin{center}
\begin{minipage}[b]{0.47\linewidth}
\includegraphics[width=\textwidth,angle=0]{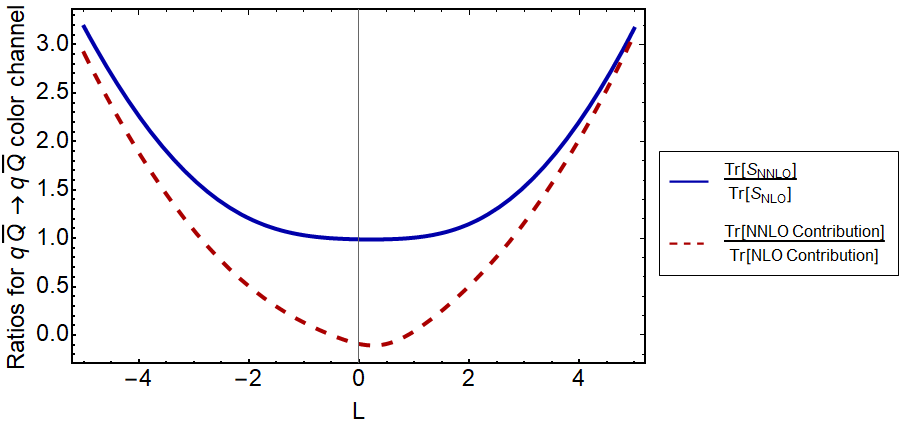}
\end{minipage}
\hspace{0.5cm}
\begin{minipage}[b]{0.47\linewidth}
\includegraphics[width=1.01\textwidth,angle=0]{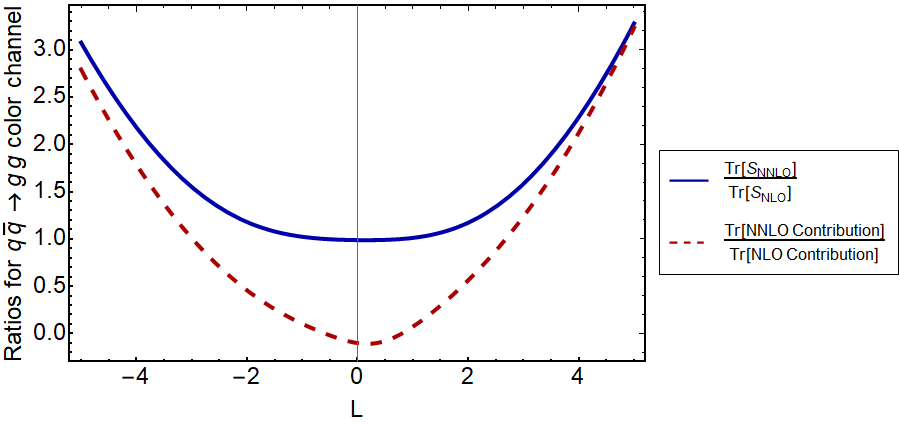}
\end{minipage}
\end{center}
\vspace{-0.5cm}
\caption{The ratios of NNLO to NLO soft function and NNLO to NLO contribution for the $q\bar{Q}\rightarrow q\bar{Q}$ and $q\bar{q}\rightarrow g g$ color channels.} \label{Ren2} 
\end{figure}

\begin{figure}[ht]
\begin{center}
\begin{minipage}[b]{0.47\linewidth}
\includegraphics[width=\textwidth,angle=0]{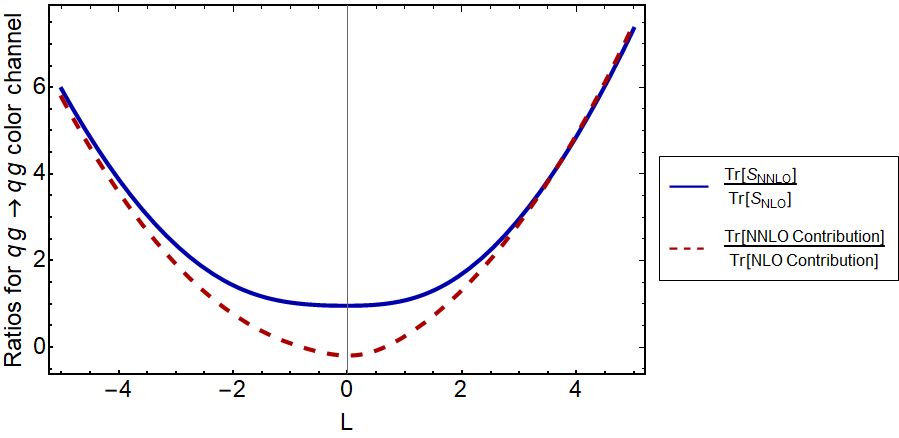}
\end{minipage}
\hspace{0.5cm}
\begin{minipage}[b]{0.47\linewidth} 
\includegraphics[width=1.01\textwidth,angle=0]{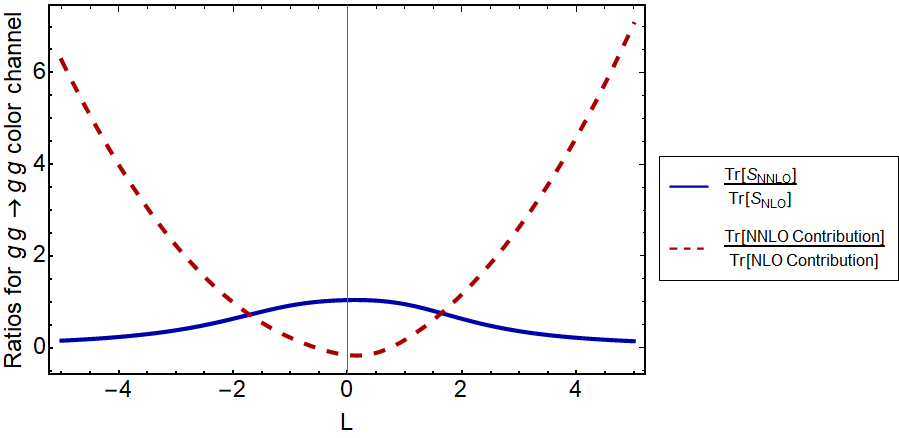}
\end{minipage}
\end{center}
\vspace{-0.5cm}
\caption{The ratios of NNLO to NLO soft function and NNLO to NLO contribution for the $q g\rightarrow q g$ and $ g g\rightarrow g g$ color channels.}\label{Ren3}  
\end{figure}

\begin{figure}[ht]
\begin{center}
\begin{minipage}[b]{0.47\linewidth}
\includegraphics[width=\textwidth,angle=0]{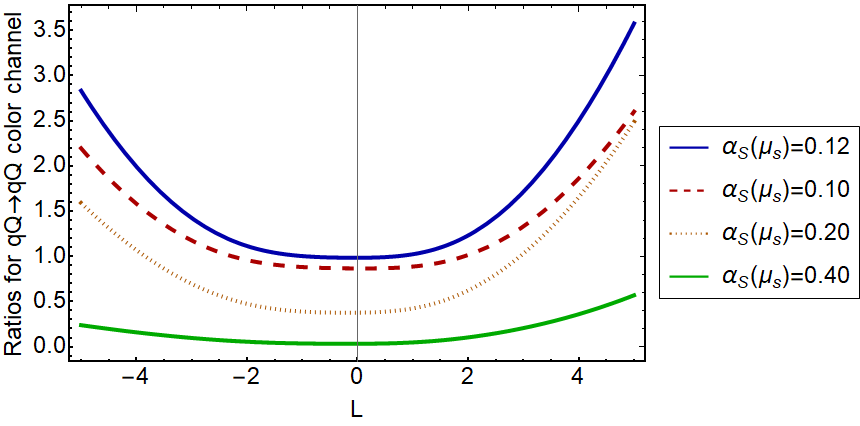}
\end{minipage}
\hspace{0.5cm}
\begin{minipage}[b]{0.47\linewidth}
\includegraphics[width=1.01\textwidth,angle=0]{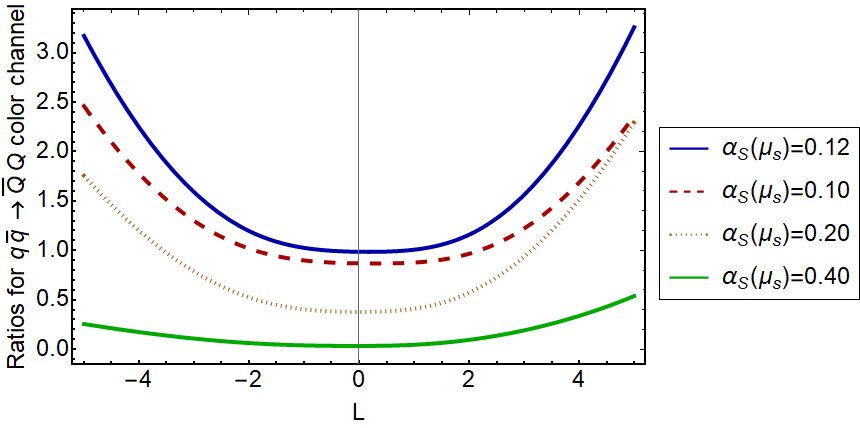}
\end{minipage}
\end{center}
\vspace{-0.5cm}
\caption{The ratios of the resummed NNLO and NLO soft function for different values of $\alpha\left(\mu_s\right)$ for $qQ\rightarrow qQ$ and $q\bar{q}\rightarrow \bar{Q}Q$ color channels.}  \label{Res1}
\end{figure}

\section{Conclusions}\label{sec:Conclusions}
In this manuscript we performed a calculation of the global soft function for single-inclusive jet production involving four light-cone directions. In comparison with the soft function involving three light-cone directions, the color structure is not diagonal which complicates the color structure. We showed by an explicit calculation that the tripole contribution proportional to $T_1\cdot T_2\cdot T_3$ vanishes.  We performed resummation for each color channel with the use of color-space formalism and show that the effects of resummation are quite significant. To give definite predictions, one has to combine the soft function which we calculated in this paper with the NNLO jet and hard functions which we refer to our future work. 

\subsubsection{Acknowledgments:} Y.B. appreciates fruitful discussions with Xiaohui Liu, Sonny Mantry, Ye Li, Thomas Becher and Guido Bell. J.G. is supported by Guangdong Innovative and Entrepreneurial Research Team Program (No. 2016ZT06G025) 

\begin{figure}[ht]
\begin{center}
\begin{minipage}[b]{0.47\linewidth}
\includegraphics[width=\textwidth,angle=0]{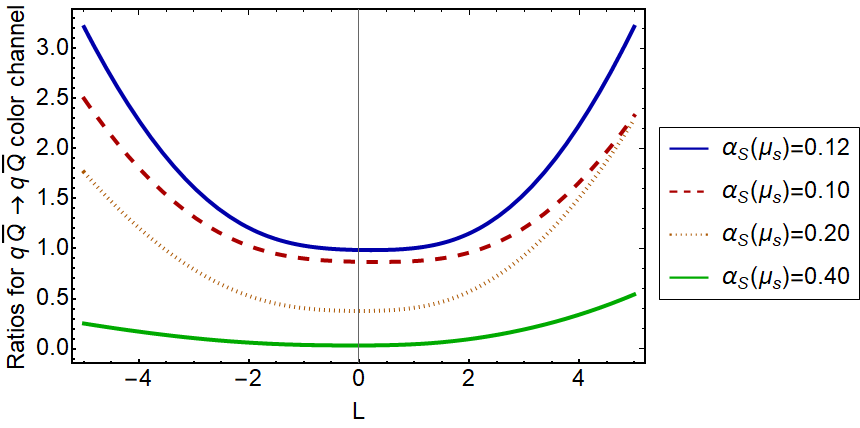}
\end{minipage}
\hspace{0.5cm}
\begin{minipage}[b]{0.47\linewidth}
\includegraphics[width=1.01\textwidth,angle=0]{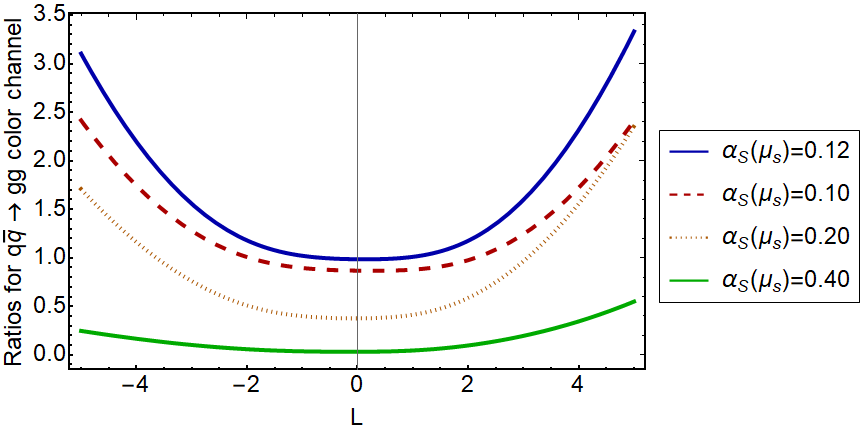}
\end{minipage}
\end{center}
\vspace{-0.5cm}
\caption{The ratios of the resummed NNLO and NLO soft function for different values of $\alpha\left(\mu_s\right)$ for $q\bar{Q}\rightarrow q\bar{Q}$ and $q\bar{q}\rightarrow g g$ color channels.} \label{fig:Res2} 
\end{figure}

\begin{figure}[ht]
\begin{center}
\begin{minipage}[b]{0.47\linewidth}
\includegraphics[width=\textwidth,angle=0]{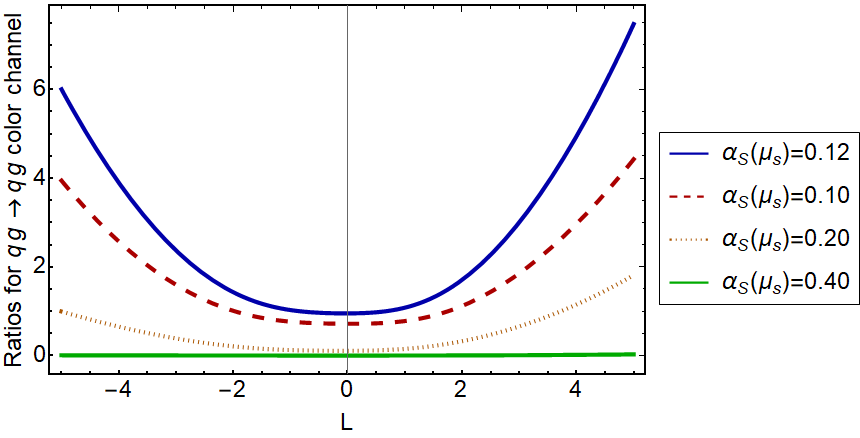}
\end{minipage}
\hspace{0.5cm}
\begin{minipage}[b]{0.47\linewidth}
\includegraphics[width=1.01\textwidth,angle=0]{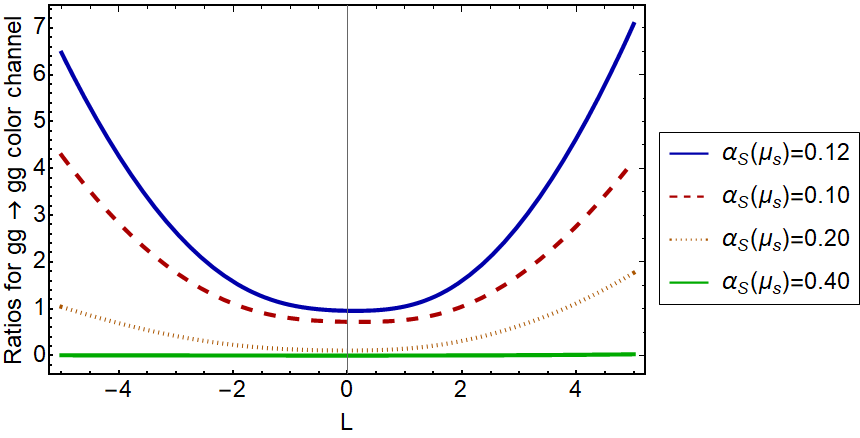}
\end{minipage}
\end{center}
\vspace{-0.5cm}
\caption{The ratios of the resummed NNLO and NLO soft function for different values of $\alpha\left(\mu_s\right)$ for $q g\rightarrow q g$ and $g g\rightarrow g g$ color channels.} \label{Res3} 
\end{figure}

\section{Auxiliary results}\label{Appendices}
\section*{Appendix A: Integrations}\label{Integrations}
\subsection*{Strongly ordered soft function}
We  consider: 
\bea
I_{1}^{s.o.} &=& \int \mathrm{d}^d q \,  \delta^{(d)}(q -  k_1 - k_2)  \, 
\int [\mathrm{d} k_1 ] [ \mathrm{d} k_2] \frac{1 }{ k_1 \cdot k_2} \, 
 \, 
\frac{  n_{ij}  }{n_i \cdot k_1 n_j \cdot k_2}   
\delta(\omega - n_4 \cdot k_1 - n_4 \cdot k_2)  \nn \\
&=& 
\int \mathrm{d}^d q \,  \frac{4 }{ q^2 } \,  \, \delta(\omega - n_4 \cdot q ) 
\int [\mathrm{d} k_1 ] [ \mathrm{d} k_2] 
 \, 
\frac{  n_{ij}  }{ 2 n_i \cdot k_1 n_j \cdot k_2}   \delta^{(d)}(q -  k_1 - k_2) 
 \nn \\
 &=& 4 \pi^{1-\epsilon}
\left[ \frac{1}{2\pi} \right]^{2d-2} \, \frac{\Gamma(-\epsilon)}{ \Gamma(1-2\epsilon) }
 \int \mathrm{d}^d q \,  \, \delta(\omega - n_4 \cdot q )  \, \nn \\
&&  \times  (q^2)^{-1-\epsilon}\, 
q_\perp^{-2-2\epsilon} \,
\left(
 \frac{2 n_i \cdot q n_j \cdot q}{n_{ij} }
 \right)^\epsilon \,  
 {}_2 F_1 \left( -\epsilon,-\epsilon, 1-\epsilon, \frac{   n_{ij} q^2 }{2 n_i \cdot q n_j \cdot q}
\right)
\eea

Now we let 
\bea
q^\mu = q_+ \frac{n_i^\mu}{ \sqrt{2 n_{ij}}}
+  q_- \frac{n_j^\mu}{ \sqrt{2 n_{ij}}}
+ q_\perp^\mu
\eea
and further 
\bea
q_+ = \frac{\omega}{n_4^-} x y  \,,  \quad \quad
q_- = \frac{\omega}{n_4^+} {\bar x} y  \,,  \quad \quad 
q_\perp = \frac{\omega}{\sqrt{   n_4^+ n_4^- }  } \sqrt{x {\bar x}}  y u\,,
\eea
to find
\bea
&& 4 \pi^{1-\epsilon} \, 
\left[ \frac{1}{2\pi} \right]^{2d-2} \, \frac{\Gamma(-\epsilon)}{ \Gamma(1-2\epsilon) }
 \frac{ \Omega_{d-3} }{2} 
 \int \mathrm{d} q_+  \, 
 \mathrm{d} q_- \,
 \mathrm{d} q_\perp  \, 
 \mathrm{d} c_\theta   \, \, 
 q_\perp^{1-2\epsilon} \, 
  s_\theta^{-1-2\epsilon}
  \, \delta(\omega - n_4 \cdot q )  \, \nn \\
&&  \times 
\left( \frac{\omega^2}{n_4^+ n_4^-}  \right)^{-2-\epsilon} 
\left( x   {\bar x} \right)^{-2-\epsilon} 
\left[ y^2 \right]^{-2-\epsilon} 
 \left(      1-u^2 \right)^{-1-\epsilon}\, 
\left(      u^2  \right)^{-1-\epsilon} \,
 \,
 {}_2 F_1 \left( -\epsilon,-\epsilon, 1-\epsilon, 1-u^2
\right) \nn \\
\eea
which is:
\bea
&& 4 \pi^{1-\epsilon} \, 
\left[ \frac{1}{2\pi} \right]^{2d-2} \, \frac{\Gamma(-\epsilon)}{ \Gamma(1-2\epsilon) }
 \frac{ \Omega_{d-3} }{2} 
 \int \mathrm{d} x  \, 
 \mathrm{d} y \,
 \mathrm{d} u  \, 
 \mathrm{d} c_\theta   \, \, 
   \left( \frac{\omega^2}{n_4^+ n_4^-}  \right)^{3/2}
(x {\bar x})^{1/2} \, 
y^2  \nn \\
&& \times   \,
 \left(   
 \frac{\omega^2}{n_4^+n_4^-} x{\bar x} y^2 u^2
  \right)^{1/2-\epsilon} \, 
  s_\theta^{-1-2\epsilon} \, 
   \frac{1}{\omega} \delta\left(1 -  
   \frac{y}{2} +  \sqrt{x{\bar x}} y u c_\theta
    \right)  \, \nn \\
&&  \times  
\left( \frac{\omega^2}{n_4^+ n_4^-}  \right)^{-2-\epsilon} 
\left( x   {\bar x} \right)^{-2-\epsilon} 
\left[ y^2 \right]^{-2-\epsilon} 
 \left(      1-u^2 \right)^{-1-\epsilon}\, 
\left(      u^2  \right)^{-1-\epsilon} \,
 \,
 {}_2 F_1 \left( -\epsilon,-\epsilon, 1-\epsilon, 1-u^2
\right) \nn \\
\eea
which simplifies to 
\bea
&& 4 \pi^{1-\epsilon} \, 
\left[ \frac{1}{2\pi} \right]^{2d-2} \, \frac{\Gamma(-\epsilon)}{ \Gamma(1-2\epsilon) }
 \frac{ \Omega_{d-3} }{2}  \,
 \frac{1}{\omega} \left( \frac{4  \, \omega^2}{n_4^+ n_4^-}  \right)^{-2 \epsilon}  \, 
 \, 
 \int \mathrm{d} x  \, 
 \mathrm{d} u  \, 
 \mathrm{d} c_\theta   \, \, 
   \nn \\
&& \times   \,
  s_\theta^{-1-2\epsilon} \,  \left(  1  - 2  \sqrt{x {\bar x} } \, u \, c_\theta   \right)^{   4\epsilon} 
  \, \nn \\
&&  \times  
\left( x   {\bar x} \right)^{-1  -2 \epsilon} 
\, 
 \left(      1-u^2 \right)^{-1-\epsilon}\, 
\left(      u^2  \right)^{-1/2 -2 \epsilon} \,
 \,
 {}_2 F_1 \left( -\epsilon,-\epsilon, 1-\epsilon, 1-u^2
\right) \nn \\
\eea
Integrating over $c_\theta$ gives
\bea
&& 4 \pi^{1-\epsilon} \, 
\left[ \frac{1}{2\pi} \right]^{2d-2} \, \frac{\Gamma(-\epsilon)}{ \Gamma(1-2\epsilon) }
 \frac{ \Omega_{d-3} }{4}  \,
 \frac{1}{\omega} \left( \frac{4  \, \omega^2}{n_4^+ n_4^-}  \right)^{-2 \epsilon}  \, 
 \, 4^{1+2\epsilon} \,
 \sqrt{\pi} \, 
\frac{ \Gamma\left(\frac{1}{2} - \epsilon \right) }{    \Gamma(1-\epsilon)    } \,   
   \nn \\
&& \times   \,
\int \mathrm{d} x  \, 
  \mathrm{d} u^2  \,     
  {}_2F_1 \left( \frac{1}{2} -2 \epsilon, -2\epsilon, 1-\epsilon, 4x{\bar x} u^2 \right)
   \, 
   \left(4  x   {\bar x} \right)^{-1  -2 \epsilon} 
   \nn \\
&&  \times  
\, 
 \left(      1-u^2 \right)^{-1-\epsilon}\, 
\left(      u^2  \right)^{-1 -2 \epsilon} \,
 \,
 {}_2 F_1 \left( -\epsilon,-\epsilon, 1-\epsilon, 1-u^2
\right) \nn \\
\eea
Now we let $4 x{\bar x} = t $ to find 
\bea
&& 4 \pi^{1-\epsilon} \, 
\left[ \frac{1}{2\pi} \right]^{2d-2} \, \frac{\Gamma(-\epsilon)}{ \Gamma(1-2\epsilon) }
 \frac{ \Omega_{d-3} }{4}  \,
 \frac{1}{\omega} \left( \frac{4  \, \omega^2}{n_4^+ n_4^-}  \right)^{-2 \epsilon}  \, 
 \, 4^{1+2\epsilon} \,
 \sqrt{\pi} \, 
\frac{ \Gamma\left(\frac{1}{2} - \epsilon \right) }{    \Gamma(1-\epsilon)    } \,   
   \nn \\
&& \times   \, \frac{1}{2}
\int \mathrm{d} t  \, 
  \mathrm{d} u^2  \,     
  {}_2F_1 \left( \frac{1}{2} -2 \epsilon, -2\epsilon, 1-\epsilon, t  u^2 \right)
   \, 
   \left( t \right)^{-1  -2 \epsilon}  (1-t)^{-1/2}
   \nn \\
&&  \times  
\, 
 \left(      1-u^2 \right)^{-1-\epsilon}\, 
\left(      u^2  \right)^{-1 -2 \epsilon} \,
 \,
 {}_2 F_1 \left( -\epsilon,-\epsilon, 1-\epsilon, 1-u^2
\right) \nn \\
\eea
which gives
\bea
&& 4 \pi^{1-\epsilon} \, 
\left[ \frac{1}{2\pi} \right]^{2d-2} \, \frac{\Gamma(-\epsilon)}{ \Gamma(1-2\epsilon) }
 \frac{ \Omega_{d-3} }{4}  \,
 \frac{1}{\omega} \left( \frac{4  \, \omega^2}{n_4^+ n_4^-}  \right)^{-2 \epsilon}  \, 
 \, 4^{1+2\epsilon} \,
 \sqrt{\pi} \, 
\frac{ \Gamma\left(\frac{1}{2} - \epsilon \right) }{    \Gamma(1-\epsilon)    } \,   
  \, \frac{1}{2}
\,   \sqrt{\pi}\, 
\frac{ \Gamma(-2\epsilon)   }{ \Gamma\left(  \frac{1}{2}  -2 \epsilon \right)}
\, 
   \nn \\
&&  \times     \int_0^1 \, \mathrm{d} u^2  \,    
\, 
 \left(      1-u^2 \right)^{-1-\epsilon}\, 
\left(      u^2  \right)^{-1 -2 \epsilon} \,
 \,
 {}_2 F_1 \left( -\epsilon,-\epsilon, 1-\epsilon, 1-u^2
\right) \,
 \,
 {}_2 F_1 \left( -2\epsilon,-2\epsilon, 1-\epsilon, u^2
\right)
 \nn \\
\eea

which is 
\bea
&& 4 \pi^{1-\epsilon} \, 
\left[ \frac{1}{2\pi} \right]^{2d-2} \, \frac{\Gamma(-\epsilon)}{ \Gamma(1-2\epsilon) }
 \frac{ 1}{4}  \, \frac{2  \pi^{1/2-\epsilon}  }{\Gamma(1/2-\epsilon)}
 \frac{1}{\omega} \left( \frac{4  \, \omega^2}{n_4^+ n_4^-}  \right)^{-2 \epsilon}  \, 
 \, 4^{1+2\epsilon} \,
 \sqrt{\pi} \, 
\frac{ \Gamma\left(\frac{1}{2} - \epsilon \right) }{    \Gamma(1-\epsilon)    } \,   
  \, \frac{1}{2}
\,   \sqrt{\pi}\, 
\frac{ \Gamma(-2\epsilon)   }{ \Gamma\left(  \frac{1}{2}  -2 \epsilon \right)}
\, 
   \nn \\
&&  \times     \int_0^1 \, \mathrm{d} u^2  \,    
\, 
 \left(      1-u^2 \right)^{-1-\epsilon}\, 
\left(      u^2  \right)^{-1 -2 \epsilon} \,{}_2 F_1 \left( -2\epsilon,-2\epsilon, 1-\epsilon, u^2
\right)
 \,
 \Bigg(
 1 + \epsilon^2  {\rm Li}_2(1-u^2)   + {\cal O}(\epsilon^3)
 \Bigg)
 \nn \\
\eea
Here we need to expand ${}_2 F_1 \left( -\epsilon,-\epsilon, 1-\epsilon, 1-u^2
\right)$ to ${\cal O}(\epsilon^3)$. The ${\cal O}(\epsilon^3)$ term is lengthy and we suppressed its expression. 

The integration gives:
\bea
I_1^{s.o.}  &= & \frac{1}{4\pi^2} \frac{c_R}{16\pi^2}  \left[
\frac{2}{\epsilon^2}
 - 2\pi^2 - \frac{112}{3} \zeta_3 \epsilon - \frac{\pi^4}{3} \epsilon^2
 \right] 
  \frac{1}{\omega} \left( \frac{4  \, \omega^2}{n_4^+ n_4^-}  \right)^{-2 \epsilon}  \, 
\, \nn \\
&& \times \left( 
- \frac{3}{2\epsilon} 
- \frac{\pi^2}{6} \epsilon 
+ 13 \zeta_3 \epsilon^2 
+ \frac{31}{180} \pi^4 \epsilon^3 
- \frac{\pi^2}{12} \epsilon  - \frac{\pi^4}{720} \epsilon^3
- \epsilon^2 \zeta_3 
\right) \nn \\
&=& \frac{1}{4\pi^2} \frac{c_R}{16\pi^2}  \, \frac{1}{\omega} \left( \frac{4  \, \omega^2}{n_4^+ n_4^-}  \right)^{-2 \epsilon}  \, 
\left(
-\frac{3}{\epsilon^3} + \frac{5\pi^2}{2\epsilon}
+ 80\zeta_3 
+ \frac{161}{120}\pi^4 \epsilon
\right)
\eea

The rest term for ${\cal S}^{s.o.}$ is given by $- (2\pi)^{2-2d} \times I_4$ in~\cite{Becher2012}, which is
\bea
I_2^{s.o.} = 
-  \frac{1}{4\pi^2} \frac{c_R }{16\pi^2}  \, \frac{1}{\omega} \left( \frac{4  \, \omega^2}{n_4^+ n_4^-}  \right)^{-2 \epsilon}  \,
 \left(
 - \frac{4}{\epsilon^3} 
 + \frac{10\pi^2}{3\epsilon} 
 + \frac{248}{3} \zeta_3 
 + \frac{25}{18} \pi^4 \epsilon 
 \right)
\eea

Therefore we have for the strongly ordered limit contribution
\bea
S_{G, ij}^{s.o.} (\omega) = 
  \frac{1}{4\pi^2} \frac{c_R }{16\pi^2}  \, \frac{1}{\omega} \left( \frac{4  \, \omega^2}{n_4^+ n_4^-}  \right)^{-2 \epsilon}  \,
 \left(
 - \frac{2}{\epsilon^3} 
 + \frac{5 \pi^2}{3\epsilon} 
 + \frac{232}{3} \zeta_3 
 + \frac{233}{180} \pi^4 \epsilon 
  \right) 
\eea

\subsection*{Calculation of $I_{7,2}$}
$ (2\pi)^{2-2d} \times I_{7,2}$ is defined as
\bea
&& \int \mathrm{d}^d q \delta(\omega - n_4 \cdot q)
\int [\mathrm{d} k_1 ] [ \mathrm{d}k_2] 
\frac{n_{ij}}{q^2} 
\frac{   n_j \cdot(k_1+2k_2)        }{ n_j\cdot q n_i \cdot k_1  n_j \cdot k_2      }
\delta^{(d)} \delta(q-k_1-k_2)  \nn \\
&=& \frac{1}{2} \, I_1^{s.o.} + \int \mathrm{d}^d q \delta(\omega - n_4 \cdot q)
\, \frac{n_{ij}}{q^2}  \,  \frac{1}{n_j\cdot q } \, 
\int [\mathrm{d} k_1 ] [ \mathrm{d}k_2]  \, 
 \,  
\frac{      1      }{ n_i \cdot k_1        }
\,
\delta^{(d)} \delta(q-k_1-k_2)  \nn \\
&=& \frac{1}{2} \, I_1^{s.o.} 
+ \pi^{1-\epsilon} \, 
\left( \frac{1}{2\pi} \right)^{2d-2} \,\frac{\Gamma(-\epsilon)}{2 \Gamma(1-2\epsilon) }  \, 
 \int \mathrm{d}^d q \delta(\omega - n_4 \cdot q)
\, \frac{n_{ij}}{(q^2)^{1+\epsilon}}  \,  \frac{1}{n_i \cdot q \, n_j\cdot q } \, 
  \nn \\
  &=&  \frac{1}{2} \, I_1^{s.o.}  
  + \pi^{1-\epsilon} \, 
\left( \frac{1}{2\pi} \right)^{2d-2} \,\frac{\Gamma(-\epsilon)}{2 \Gamma(1-2\epsilon) }  \, 
I_2 \nn \\
&=&
\frac{1}{(2\pi)^2}
 \frac{c_R}{16\pi^2} 
 \frac{1}{\omega} \left( \frac{4  \, \omega^2}{n_4^+ n_4^-}  \right)^{-2 \epsilon} \,
\left(
- \frac{2}{\epsilon^3} + \frac{3\pi^2}{2\epsilon} 
+ \frac{151}{3}\zeta_3 \, 
+ \frac{11}{12}\pi^4 \epsilon
\right)
\eea

\subsection*{Calculation of $I_6$}
$ (2\pi)^{2-2d} \times I_{6}$ is defined as
\bea
&& \int \mathrm{d}^d q \delta(\omega - n_4 \cdot q)
\frac{n_{ij}}{q^2} 
\int [\mathrm{d} k_1 ] [ \mathrm{d}k_2] 
\frac{   n_j \cdot(k_1- k_2)        }{ n_i\cdot q n_j \cdot q n_j\cdot k_2    }
\delta^{(d)} \delta(q-k_1-k_2)  \nn \\
&=& \int \mathrm{d}^d q \delta(\omega - n_4 \cdot q)
\int [\mathrm{d} k_1 ] [ \mathrm{d}k_2] 
\left( \frac{n_{ij}}{q^2} 
\frac{  1     }{ n_i\cdot q   n_j\cdot k_2    }
- \frac{n_{ij}}{q^2} 
\frac{     2         }{ n_i\cdot q n_j \cdot q     }
\right)
\delta^{(d)} \delta(q-k_1-k_2)  \nn \\
&=& \int \mathrm{d}^d q \delta(\omega - n_4 \cdot q)
\frac{n_{ij}}{q^2}  \, \frac{  1     }{ n_i\cdot q  }
\int [\mathrm{d} k_1 ] [ \mathrm{d}k_2] 
\frac{  1     }{   n_j\cdot k_2    }
\delta^{(d)} \delta(q-k_1-k_2)  \nn \\
&& - 
 \int \mathrm{d}^d q \delta(\omega - n_4 \cdot q)
\,
 \frac{n_{ij}}{q^2} 
\frac{     2         }{ n_i\cdot q n_j \cdot q     }
\,
\int [\mathrm{d} k_1 ] [ \mathrm{d}k_2] 
\delta^{(d)} \delta(q-k_1-k_2)
 \nn \\
 &=& 
\left( \frac{1}{2\pi} \right)^{2d-2} \,  \frac{ \pi^{1-\epsilon} }{2}  \,
\left[ 
\frac{\Gamma(-\epsilon)}{\Gamma(1-2\epsilon) }
- 
\frac{2 \, \Gamma(1-\epsilon)}{\Gamma(2-2\epsilon)}  
\right] 
\,
 \int \mathrm{d}^d q \delta(\omega - n_4 \cdot q)
\frac{n_{ij}}{(q^2)^{1+\epsilon}}  \, \frac{  1     }{ n_i\cdot q n_j \cdot q }
 \nn \\
  &=& 
\left( \frac{1}{2\pi} \right)^{2d-2} \,  \frac{ \pi^{1-\epsilon} }{2}  \,
\left[ 
\frac{\Gamma(-\epsilon)}{\Gamma(1-2\epsilon) }
- 
\frac{2 \, \Gamma(1-\epsilon)}{\Gamma(2-2\epsilon)}  
\right] 
\,
I_2 \nn \\
&=& 
\left( \frac{1}{2\pi} \right)^{2d-2} \,   
\frac{ - \pi \csc(2\epsilon \pi) \Gamma^2(-\epsilon)      }{(1-2\epsilon) \Gamma(1-4\epsilon)}
\,
\,\frac{1}{\omega} \left( \frac{4  \, \omega^2}{n_4^+ n_4^-}  \right)^{-2 \epsilon} 
\eea

\section*{Appendix B: Color structure}
\label{Color}

For the case of 4 or more partons involved, the color structure is not diagonal anymore like in~\cite{Becher2012} since the color conservation relation alone is not sufficient to express all products $T_i \cdot T_j$ in terms of ${T}_i^2$. Therefore, we calculate the action of $T_i \cdot T_j$ on the color basis and the matrices
$(\omega_{ij})_{IJ} = \langle \theta_I| T_i \cdot T_j | \theta_J\rangle$. 
A final color matrix $(\omega_{ij})_{IJ}$is a   matrix whose \textit{columns} are action of $\vec{T}_i\cdot \vec{T}_j$ 
on basis vectors and multiply it 
from the left on the tree level 
color matrix. It is convenient to calculate it in a form: $\omega = \langle \theta| A | \theta\rangle = \underbrace{[\theta^H \theta]}_{Tree \ level} [A^T]$, where $\underbrace{[\theta^H \theta]}_{Tree \ level}$ - tree level color matrix. We consider each channel separately. Here $N_c = 3$ represents the number of colors, $C_F = \frac{N_c^2 - 1}{2 N_c}$, and $C_A = N_c$.

\subsection*{Channel $q_1Q_2\rightarrow q_3 Q_4$}

Our color basis follows the conventions of~\cite{Color_qQqQ_1, Color_qQqQ_2}: 
$$
\begin{cases} 
|C_1\rangle = (\bar{q}_{i_4}\delta_{i_4 i_1} q_{i_1})(\bar{q}_{i_3}\delta_{i_3 i_2} q_{i_2}) \\ 
|C_2\rangle = (\bar{q}_{i_4}\delta_{i_4 i_2} q_{i_2})(\bar{q}_{i_3}\delta_{i_3 i_1} q_{i_1})
\end{cases};
\quad
[\theta^T\theta]=
\begin{pmatrix} 
N_c^2 & N_c \\
N_c & N_c^2 
\end{pmatrix}
$$

$$ T_1\cdot T_2=
\begin{pmatrix} 
-\frac{1}{2N_c} & \frac{1}{2} \\
\frac{1}{2} & -\frac{1}{2N_c} 
\end{pmatrix};
\quad
T_1\cdot T_3=
\begin{pmatrix} 
\frac{1}{2N_c} & 0 \\
-\frac{1}{2} & -C_F 
\end{pmatrix};
\quad
T_1\cdot T_4=
\begin{pmatrix} 
-C_F & -\frac{1}{2} \\
0 & \frac{1}{2N_c} 
\end{pmatrix}
$$ 

$$
\omega_{12}=
(\frac{N^2_c-1}{2})\cdot
\begin{pmatrix} 
0 & 1 \\
1 & 0 
\end{pmatrix};
\quad
\omega_{13}=
(\frac{1-N^2_c}{2})\cdot
\begin{pmatrix} 
0 & 1 \\
1 & N_c 
\end{pmatrix};
\quad
\omega_{14}=
(\frac{1-N^2_c}{2})\cdot
\begin{pmatrix} 
N_c & 1 \\
1 & 0 
\end{pmatrix}
$$

By color conservation: 
$\omega_{23}= \omega_{14}$, 
$\omega_{24} = \omega_{13}$, 
$\omega_{34} = \omega_{12}$.

\subsection*{Channel $q_1\bar{q}_2\rightarrow \bar{Q}_3Q_4$}
Our color basis follows the conventions of~\cite{Color_qQqQ_1, Color_qQqQ_2}: 
$$
\begin{cases} 
|C_1\rangle = (\bar{q}_{i_4}\delta_{i_4 i_1} q_{i_1})(\bar{q}_{i_2}\delta_{i_2 i_3} q_{i_3}) \\ 
|C_2\rangle = (\bar{q}_{i_4}\delta_{i_4 i_3} q_{i_3})(\bar{q}_{i_2}\delta_{i_2 i_1} q_{i_1})
\end{cases};
\quad
[\theta^T\theta]=
\begin{pmatrix} 
N_c^2 & N_c \\
N_c & N_c^2 
\end{pmatrix}
$$

$$ T_1\cdot T_2=
\begin{pmatrix} 
\frac{1}{2N_c} & 0 \\
-\frac{1}{2} & -C_F 
\end{pmatrix};
\quad
T_1\cdot T_3=
\begin{pmatrix} 
-\frac{1}{2N_c} & \frac{1}{2} \\
\frac{1}{2} & -\frac{1}{2N_c} 
\end{pmatrix};
\quad
T_1\cdot T_4=
\begin{pmatrix} 
-C_F & -\frac{1}{2} \\
0 & \frac{1}{2N_c} 
\end{pmatrix}
$$

$$
\omega_{12}=
(\frac{1 - N^2_c}{2})\cdot
\begin{pmatrix} 
0 & 1 \\
1 & N_c 
\end{pmatrix};
\quad
\omega_{13}=
(\frac{N^2_c - 1}{2})\cdot
\begin{pmatrix} 
0 & 1 \\
1 & 0 
\end{pmatrix};
\quad
\omega_{14}=
(\frac{1-N^2_c}{2})\cdot
\begin{pmatrix} 
N_c & 1 \\
1 & 0 
\end{pmatrix}
$$

By color conservation:  $\omega_{23} = \omega_{14}$, $\omega_{24} = \omega_{13},$
$\omega_{34} = \omega_{12}$.

\subsection*{Channel $q_1\bar{Q}_2\rightarrow q_3\bar{Q}_4$}
Our color basis follows the conventions of~\cite{Color_qQqQ_1, Color_qQqQ_2}: 

$$
\begin{cases} 
|C_1\rangle = (\bar{q}_{i_2}\delta_{i_2 i_1} q_{i_1})(\bar{q}_{i_3}\delta_{i_3 i_4} q_{i_4}) \\ 
|C_2\rangle = (\bar{q}_{i_2}\delta_{i_2 i_4} q_{i_4})(\bar{q}_{i_3}\delta_{i_3 i_1} q_{i_1})
\end{cases};
\quad
[\theta^T\theta]=
\begin{pmatrix} 
N_c^2 & N_c \\
N_c & N_c^2 
\end{pmatrix}
$$

$$
T_1\cdot T_2=
\begin{pmatrix} 
-C_F & -\frac{1}{2} \\
0 & \frac{1}{2N_c} 
\end{pmatrix};
\quad
T_1\cdot T_3=
\begin{pmatrix} 
\frac{1}{2N_c} & 0 \\
-\frac{1}{2} & -C_F
\end{pmatrix};
\quad
T_1\cdot T_4=
\begin{pmatrix} 
-\frac{1}{2N_c} & \frac{1}{2} \\
\frac{1}{2} & -\frac{1}{2N_c} 
\end{pmatrix}
$$

$$
\omega_{12}=
(\frac{1 - N^2_c}{2})\cdot
\begin{pmatrix} 
N_c & 1 \\
1 & 0 
\end{pmatrix};
\quad
\omega_{13}=
(\frac{1 - N^2_c}{2})\cdot
\begin{pmatrix} 
0 & 1 \\
1 & N_c 
\end{pmatrix};
\quad
\omega_{14}=
(\frac{N^2_c - 1}{2})\cdot
\begin{pmatrix} 
0 & 1 \\
1 & 0 
\end{pmatrix}
$$

By color conservation:  $\omega_{23} = \omega_{14}$,
$\omega_{24} = \omega_{13}$
, $\omega_{34} = \omega_{12}$.

\subsection*{Channel $q_1\bar{q}_2\rightarrow g_3 g_4$}

For convenience, for the channels with 
2 (anti)quarks and 2 gluons, we rescale the Gell-Mann matrices, however, leaving the color operators the same.

Our conventions follow~\cite{Color_quark_gluon1}:

$$
tr(t^at^b) = \delta^{ab},\ tr(T^aT^b) = \frac{\delta^{ab}}{2}\Rightarrow t^a = \sqrt{2} T^a
$$ 

Color basis and tree-level color matrix are: 
$$
\begin{cases} 
|C_1\rangle = (\bar{\xi}_{i_2}(t^{a_3}t^{a_4})_{i_2 i_1}\xi_{i_1})A_{a_3}A_{a_4} \\ 
|C_2\rangle = (\bar{\xi}_{i_2}(t^{a_4}t^{a_3})_{i_2 i_1}\xi_{i_1})A_{a_3}A_{a_4} \\
|C_3\rangle = (\bar{\xi}_{i_2}\delta^{a_3 a_4}\delta_{i_2 i_1}\xi_{i_1})A_{a_3}A_{a_4}
\end{cases};
\quad
[\theta^T\theta]=
(\frac{N_c^2-1}{N_c})\cdot
\begin{pmatrix} 
N_c^2-1 & -1 & N_c \\
-1  & N_c^2-1 & N_c \\ 
N_c & N_c    & N_c^2 
\end{pmatrix}
$$

$$
T_1\cdot T_2=
\begin{pmatrix} 
\frac{1}{2N_c} & 0 & 0 \\
0 & \frac{1}{2N_c} & 0 \\
-\frac{1}{2} & -\frac{1}{2} & -C_F
\end{pmatrix};
\quad
T_1\cdot T_3=
\begin{pmatrix} 
0 & 0 & \frac{1}{2} \\
0 & -\frac{N_c}{2} & -\frac{1}{2} \\
\frac{1}{2} & 0 & 0
\end{pmatrix};
\quad
T_1\cdot T_4=
\begin{pmatrix} 
-\frac{N_c}{2} & 0 & -\frac{1}{2} \\
0 & 0 & \frac{1}{2} \\
0 & \frac{1}{2} & 0
\end{pmatrix}
$$
$$
T_3\cdot T_4=
\begin{pmatrix} 
-\frac{N_c}{2} & 0 & 0 \\
0 & -\frac{N_c}{2} & 0 \\
-\frac{1}{2} & -\frac{1}{2} & -N_c
\end{pmatrix}
$$ 

$$
\omega_{12}=
(\frac{N^2_c-1}{N_c})\cdot
\begin{pmatrix} 
-\frac{1}{2N_c} & -\frac{N^2_c+1}{2N_c} & \frac{1}{2}(1-N^2_c) \\
-\frac{N^2_c+1}{2N_c} & -\frac{1}{2N_c} & \frac{1}{2}(1-N^2_c) \\
\frac{1}{2}(1-N^2_c) & \frac{1}{2}(1-N^2_c) & \frac{N_c}{2}(1-N^2_c)
\end{pmatrix};
$$
$$
\omega_{13}=
(\frac{N^2_c-1}{2})\cdot
\begin{pmatrix} 
1 & 1 & N_c \\
1 & (1-N^2_c) & -N_c \\
N_c & -N_c & 0
\end{pmatrix};
\quad
\omega_{14}=
(\frac{N^2_c-1}{2})\cdot
\begin{pmatrix} 
1-N^2_c & 1 & -N_c \\
1 & 1 & N_c \\
-N_c & N_c & 0
\end{pmatrix}
$$

$$
\omega_{34}=
N_c(1-N^2_c)\cdot
\begin{pmatrix} 
\frac{N_c}{2} & 0 & 1 \\
0 & \frac{N_c}{2} & 1 \\
1 & 1 & N_c
\end{pmatrix}
$$ 

By color conservation, $\omega_{24} = \omega_{13}$, $\omega_{23} = \omega_{14}$.

\subsection*{Channel $q_1 g_2\rightarrow g_3 q_4$}
For convenience, for the channels with 
2 (anti)quarks and 2 gluons, we rescale the Gell-Mann matrices, however, leaving the color operators the same.

Our conventions follow~\cite{Color_quark_gluon1}:
$$
\begin{cases} 
|C_1\rangle = (\bar{\xi}_{i_4}(t^{a_3}t^{a_2})_{i_4 i_1}\xi_{i_1})A_{a_3}A_{a_2} \\ 
|C_2\rangle = (\bar{\xi}_{i_4}(t^{a_2}t^{a_3})_{i_4 i_1}\xi_{i_1})A_{a_3}A_{a_2} \\
|C_3\rangle = (\bar{\xi}_{i_4}\delta^{a_2 a_3}\delta_{i_4 i_1}\xi_{i_1})A_{a_3}A_{a_4}
\end{cases};
\quad
[\theta^T\theta]=
(\frac{N_c^2-1}{N_c})\cdot
\begin{pmatrix} 
N_c^2-1 & -1 & N_c \\
-1  & N_c^2-1 & N_c \\ 
N_c & N_c    & N_c^2 
\end{pmatrix}
$$

$$
T_1\cdot T_2=
\begin{pmatrix} 
-\frac{N_c}{2} & 0 & -\frac{1}{2} \\
0 & 0 & \frac{1}{2} \\
0 & \frac{1}{2} & 0
\end{pmatrix};
\quad
T_1\cdot T_3=
\begin{pmatrix} 
0 & 0 & \frac{1}{2} \\
0 & -\frac{N_c}{2} & -\frac{1}{2} \\
\frac{1}{2} & 0 & 0
\end{pmatrix};
\quad
T_1\cdot T_4=
\begin{pmatrix} 
\frac{1}{2N_c} & 0 & 0 \\
0 &  \frac{1}{2N_c} & 0  \\
-\frac{1}{2} & -\frac{1}{2} & -C_F
\end{pmatrix}
$$
$$
T_2\cdot T_3=
\begin{pmatrix} 
-\frac{N_c}{2} & 0 & 0 \\
0 & -\frac{N_c}{2} & 0 \\
-\frac{1}{2} & -\frac{1}{2} & -N_c
\end{pmatrix}
$$ 

$$
\omega_{12}=
(\frac{N_c^2-1}{2})\cdot
\begin{pmatrix} 
1-N^2_c & 1 & -N_c \\
1 & 1 & N_c \\
-N_c & N_c & 0
\end{pmatrix};
\quad
\omega_{13}=
(\frac{N_c^2-1}{2})\cdot
\begin{pmatrix} 
1 & 1 & N_c \\
1 & 1-N_c^2 & -N_c \\
N_c & -N_c & 0
\end{pmatrix}
$$
$$
\omega_{14}=
(\frac{1-N_c^2}{2N_c})\cdot
\begin{pmatrix} 
\frac{1}{N_c} & \frac{N_c^2+1}{N_c} & N_c^2 - 1 \\
\frac{N_c^2+1}{N_c} & \frac{1}{N_c} &  N_c^2 - 1 \\
N_c^2 - 1 & N_c^2 - 1 & N_c(N_c^2 - 1)
\end{pmatrix};
\quad
\omega_{23}=
N_c(1-N_c^2)\cdot
\begin{pmatrix} 
\frac{N_c}{2} & 0 & 1 \\
0 & \frac{N_c}{2} &  1 \\
1 & 1 & N_c
\end{pmatrix}
$$

\subsection*{Channel $gg\rightarrow gg$} 
Our conventions follow~\cite{Color_gluons}:
$$
\begin{cases} 
\mathcal{C}_1 =  tr(t^{a_1}t^{a_2}t^{a_3}t^{a_4})A^{a_1}A^{a_2}A^{a_3}A^{a_4}\\ 
\mathcal{C}_2 =  tr(t^{a_1}t^{a_2}t^{a_4}t^{a_3})A^{a_1}A^{a_2}A^{a_3}A^{a_4}\\ 
\mathcal{C}_3 =  tr(t^{a_1}t^{a_4}t^{a_2}t^{a_3})A^{a_1}A^{a_2}A^{a_3}A^{a_4}\\ 
\mathcal{C}_4 =  tr(t^{a_1}t^{a_3}t^{a_2}t^{a_4})A^{a_1}A^{a_2}A^{a_3}A^{a_4}\\ 
\mathcal{C}_5 =  tr(t^{a_1}t^{a_3}t^{a_4}t^{a_2})A^{a_1}A^{a_2}A^{a_3}A^{a_4}\\ 
\mathcal{C}_6 =  tr(t^{a_1}t^{a_4}t^{a_3}t^{a_2})A^{a_1}A^{a_2}A^{a_3}A^{a_4}\\ 
\mathcal{C}_7 =  tr(t^{a_1}t^{a_2})tr(t^{a_3}t^{a_4})A^{a_1}A^{a_2}A^{a_3}A^{a_4}\\ 
\mathcal{C}_8 =  tr(t^{a_1}t^{a_3})tr(t^{a_2}t^{a_4})A^{a_1}A^{a_2}A^{a_3}A^{a_4}\\ 
\mathcal{C}_9 =  tr(t^{a_1}t^{a_4})tr(t^{a_2}t^{a_3})A^{a_1}A^{a_2}A^{a_3}A^{a_4} 
\end{cases};
$$

Tree-level matrix in our basis is: 

$$
[\mathcal{C}^T\mathcal{C}] = 
\frac{C_F}{8C_A}
\begin{pmatrix} 
a &	b &	b &	b &	b &	c &	d &	-e & d \\
b &	a &	b &	b &	c &	b &	d &	d &	-e \\
b &	b &	a &	c &	b &	b &	-e & d & d \\
b &	b &	c &	a &	b &	b &	-e & d & d \\
b &	c &	b &	b &	a &	b &	d &	d &	-e \\
c &	b &	b &	b &	b &	a &	d &	-e & d \\
d &	d &	-e & -e & d & d & d \cdot e &e^2 &e^2 \\
-e & d & d & d & d & -e & e^2 &	d \cdot e &	e^2 \\
d & -e & d & d & -e & d & e^2 & e^2 & d \cdot e 
\end{pmatrix};
$$
with the conventions:
$$
\begin{cases} 
a = C^4_A - 3 C^2_A + 3 \\ 
b = 3 - C_A^2 \\
c = 3 +C_A^2  \\
d = 2 C^2_A C_F \\
e = C_A
\end{cases};
$$
$$
(T_1\cdot T_2) = 
\begin{pmatrix} 
-\frac{C_A}{2} & 0 & 0 & 0 & 0 & 0 & 0 & 0 & -\frac{1}{2} \\
0 & -\frac{C_A}{2} & 0 & 0 & 0 & 0 & 0 & -\frac{1}{2} & 0 \\
0 & 0 & 0 & 0 & 0 & 0 & 0 & \frac{1}{2} & \frac{1}{2} \\
0 & 0 & 0 & 0 & 0 & 0 & 0 & \frac{1}{2} & \frac{1}{2} \\
0 & 0 & 0 & 0 & -\frac{C_A}{2} & 0 & 0 & -\frac{1}{2} & 0 \\
0 & 0 & 0 & 0 & 0 & -\frac{C_A}{2} & 0 & 0 & -\frac{1}{2} \\
-\frac{1}{2} & -\frac{1}{2} & 0 & 0 & -\frac{1}{2} & -\frac{1}{2} & -C_A & 0 & 0 \\
0 & 0 & \frac{1}{2} & \frac{1}{2} & 0 & 0 & 0 & 0 & 0 \\
0 & 0 & \frac{1}{2} & \frac{1}{2} & 0 & 0 & 0 & 0 & 0 
\end{pmatrix};
$$

$$
(T_1\cdot T_3) = 
\begin{pmatrix} 
0 & 0 & 0 & 0 & 0 & 0 & \frac{1}{2} & 0 & \frac{1}{2} \\
0 & -\frac{C_A}{2} & 0 & 0 & 0 & 0 & -\frac{1}{2} & 0 & 0 \\
0 & 0 & -\frac{C_A}{2} & 0 & 0 & 0 & 0 & 0 & -\frac{1}{2} \\
0 & 0 & 0 & -\frac{C_A}{2} & 0 & 0 & 0 & 0 & -\frac{1}{2} \\
0 & 0 & 0 & 0 & -\frac{C_A}{2} & 0 & -\frac{1}{2}  & 0 & 0 \\
0 & 0 & 0 & 0 & 0 & 0 & \frac{1}{2} & 0 & \frac{1}{2} \\
\frac{1}{2} & 0 & 0 & 0 & 0 & \frac{1}{2} & 0 & 0 & 0 \\
0 & -\frac{1}{2} & -\frac{1}{2} & -\frac{1}{2} & -\frac{1}{2} & 0 & 0 & -C_A & 0 \\
\frac{1}{2} & 0 & 0 & 0 & 0 & \frac{1}{2} & 0 & 0 & 0 
\end{pmatrix};
$$

$$
(T_1\cdot T_4) = 
\begin{pmatrix} 
-\frac{C_A}{2} & 0 & 0 & 0 & 0 & 0 & -\frac{1}{2} & 0 & 0 \\
0 & 0 & 0 & 0 & 0 & 0 &\frac{1}{2} & \frac{1}{2} & 0 \\
0 & 0 & -\frac{C_A}{2} & 0 & 0 & 0 & 0 & -\frac{1}{2} & 0 \\
0 & 0 & 0 & -\frac{C_A}{2} & 0 & 0 & 0 &  -\frac{1}{2} & 0 \\
0 & 0 & 0 & 0 & 0 & 0 & \frac{1}{2} & \frac{1}{2} & 0 \\
0 & 0 & 0 & 0 & 0 & -\frac{C_A}{2} & -\frac{1}{2} & 0 & 0 \\
0 & \frac{1}{2}& 0 & 0 & \frac{1}{2} & 0 & 0 & 0 & 0 \\
0 & \frac{1}{2}& 0 & 0 & \frac{1}{2} & 0 & 0 & 0 & 0  \\
-\frac{1}{2} & 0 & -\frac{1}{2} & -\frac{1}{2} & 0 & -\frac{1}{2} & 0 & 0 & -C_A
\end{pmatrix};
$$

Using the following conventions:
$$
\begin{cases} 
t = N^2_c-1 \\
u = -\frac{N^4_c}{2}+N^2_c-1 \\ 
v = \frac{N^2_c-2}{2} \\
x = -N^2_c-1  \\
y = N\cdot(1-N^2_c) \\
z = -\frac{N^3_c}{2}
\end{cases};
$$

$\omega$-matrices are expressed as:
$$\omega_{12} = 
\frac{C_F}{8}\cdot
\begin{pmatrix} 
u & -1 & v & v & -1 & x & y & 0 & z \\
-1 & u & v & v & x & -1 & y & z & 0 \\
v & v & t & t & v & v & N_c & -z & -z \\
v & v & t & t & v & v & N_c & -z & -z \\
-1 & x & v & v & u & -1 & y & z & 0 \\
x & -1 & v & v & -1 & u & y & 0 &z \\
y & y & N_c & N_c & y & y & N_c \cdot y & -N^2_c & -N^2_c \\
0 & z & -z & -z & z & 0 & -N^2_c & 0 & N^2_c \\
z & 0 & -z & -z & 0 & z & -N^2_c & N^2_c & 0
\end{pmatrix}
$$

$$\omega_{13} = 
\frac{C_F}{8}\cdot
\begin{pmatrix} 
t & v & v & v & v & t & -z & N_c & -z \\
v & u & -1 & -1 & x & v & z & y & 0 \\
v & -1 & u & x & -1 & v & 0 & y & z \\
v & -1 & x & u & -1 & v & 0 & y & z \\
v & x & -1 & -1 & u & v & z & y & 0 \\
t & v & v & v & v & t & -z & N_c & -z \\
-z & z & 0 & 0 & z & -z & 0 & -N^2_c & N^2_c \\
N_c & y & y & y & y & N_c & -N^2_c & N_c\cdot y & -N^2_c \\
-z & 0 & z & z & 0 & -z & N^2_c & -N^2_c & 0
\end{pmatrix}
$$

$$\omega_{14} = 
\frac{C_F}{8}\cdot
\begin{pmatrix} 
u & v & -1 & -1 & v & x & z & 0 & y \\
v & t & v & v & t & v & -z & -z & N_c \\
-1 & v & u & x & v & -1 & 0 & z & y \\
-1 & v & x & u & v & -1 & 0 & z & y \\
v & t & v & v & t & v & -z & -z & N_c \\
x & v & -1 & -1 & v & u & z & 0 & y \\
z & -z & 0 & 0 & -z & z & 0 & N^2_c & -N^2_c 
\\
0 & -z & z & z & -z & 0 & N^2_c & 0 & -N^2_c\\
y & N_c & y & y & N_c & y & -N^2_c & -N^2_c & N_c\cdot y 
\end{pmatrix}
$$

\end{document}